\documentclass{aa}  

\usepackage{graphicx}
\usepackage{txfonts}
\usepackage{amsmath}
\usepackage[colorlinks=true, citecolor = blue]{hyperref}
\usepackage{subfigure}
\begin{document} 

   \title{Turbulent processing of PAHs in protoplanetary discs}
   \titlerunning{Turbulent processing of PAHs in protoplanetary discs} 
   \subtitle{Coagulation and freeze-out leading to depletion of gas-phase PAHs}

   \author{K. Lange\inst{1},
          C. Dominik\inst{1}
          \and A. G. G. M. Tielens \inst{2,}\inst{3}}

   \institute{Anton Pannekoek Institute for Astronomy, University of Amsterdam,
              Science-Park 904, 1098 XH Amsterdam, Netherlands;
              \email{k.lange@uva.nl}
         \and
             {Leiden Observatory, Leiden University, P.O. Box 9513, 2300 RA Leiden, Netherlands}
        \and {Astronomy Department, University of Maryland, MD 20742, USA}
             }

   \date{Received xxx; accepted xxx}

  \abstract
{Polycyclic aromatic hydrocarbons (PAHs) have been detected in numerous circumstellar discs. Despite the correlation between stellar temperature and low PAH detections rates, the diversity of PAH detections and non-detections at similar stellar properties is not well understood.}
    {We propose the continuous processing of PAHs through clustering, adsorption on dust grains, and their reverse-processes as key mechanisms to reduce the emission-capable PAH abundance in protoplanetary discs. This cycle of processing is driven by vertical turbulence in the disc mixing PAHs between the disc midplane and the photosphere.}
    {We used a theoretical Monte Carlo model for photodesorption in the photosphere and a coagulation code in the disc midplane to estimate the relevance and timescale of these processes in a Herbig Ae/Be disc environment. By combining these components in a 1D vertical model, we calculated the gas-phase depletion of PAHs that stick as clusters on dust grains.}
    {Our results show that the clustering of gas-phase PAHs is very efficient, and that clusters with more than 100 monomers can grow for years before they are able to freeze out in the disc midplane. Once a PAH cluster is frozen on the dust grain surface, the large heat capacity of these clusters prevents them from evaporating off the grains in UV-rich environments such as the photosphere.
    Therefore, the clustering of PAHs followed by freeze-out can lead to a depletion of gas-phase PAHs in protoplanetary discs. We find that this mechanism is more efficient when the   PAH species has fewer carbon atoms. In contrast, PAH monomers and very small clusters consisting of a few monomers can easily detach from the grain by absorption of a single UV photon. 
    Evaluated over the lifetime of protoplanetary discs, we find a depletion of PAHs by a factor that ranges between 50 and 1000 compared to the standard ISM abundance of PAHs in the inner disc through turbulent processing.}
    {Through these processes, we favour PAHs smaller than circumovalene (C$_{66}$H$_{20}$) as the major gas-phase emitters of the disc photosphere as larger PAH monomers cannot photodesorb from the grain surface. These gas-phase PAHs   co-exist  with large PAH clusters sticking on dust grains. We find a close relation between the amount of PAHs frozen out on dust grains and the dust population, as well as the strength of the vertical turbulence.}

   \keywords{protoplanetary discs - astrochemistry - stars: variables: Herbig Ae/Be}

   \maketitle

%%%%%%%%%%%%%%%%%%%%%%%%%%%%%%%%%%%%%%%%%%%%%%%%%%%%%%%%%%%%%%%%%%%%%%%%%%%%%%
%%%%%%%%%%%%%%%%%%%%%%%%%%%%%%%%%%%%%%%%%%%%%%%%%%%%%%%%%%%%%%%%%%%%%%%%%%%%%%
%%%%%%%%%%%%%%%%%%%%%%%%%%%%%%%%%%%%%%%%%%%%%%%%%%%%%%%%%%%%%%%%%%%%%%%%%%%%%%
%%%%%%%%%%%%%%%%%%%%%%%%%%%%%%%%%%%%%%%%%%%%%%%%%%%%%%%%%%%%%%%%%%%%%%%%%%%%%%
%%%%%%%%%%%%%%%%%%%%%%%%%%%%%%%%%%%%%%%%%%%%%%%%%%%%%%%%%%%%%%%%%%%%%%%%%%%%%%
\begin{figure*}
    \centering
    \includegraphics[width=0.75\linewidth]{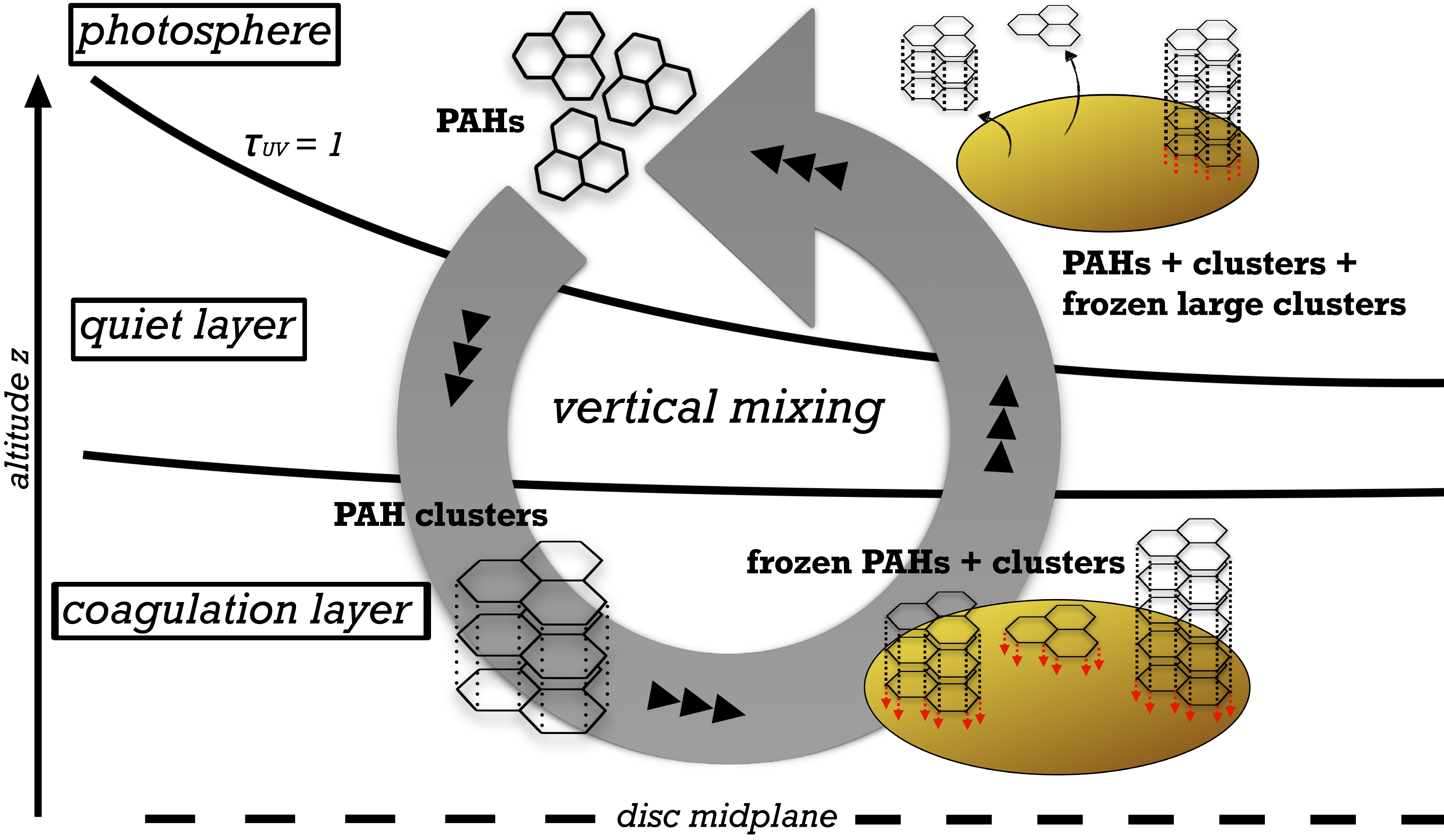}
    \caption{Sketch of PAH processing cycle in the protoplanetary disc driven by vertical mixing. PAH monomers will eventually mix into the UV depleted coagulation layer where PAH collisions lead to sticking. The cluster growth continues until the PAH clusters freeze out on dust grains. If the cluster carrying grains are mixed into the UV-rich photosphere, PAH monomers and small clusters can desorb while large clusters remain on the grain. Every cycle of coagulation, adsorption, desorption, and dissociation leads to a depletion of gas-phase PAHs as with every cycle large clusters   form and remain on the grains.}
    \label{fig:sketch}
\end{figure*}

\section{Introduction}
The signatures of polycyclic aromatic hydrocarbons (PAHs) have been identified in space through the detection of prominent mid-infrared (MIR) emission features located at 3.3, 6.2, 7.7, 8.6, and 11.2\,$\mu$m \citep{Allamandola1985, Allamandola1989}.
In protoplanetary discs, given their small size, PAHs are well coupled to the gas and are therefore present in the disc's photosphere. 
There, PAHs efficiently absorb UV radiation and convert the energy to MIR radiation, which makes them an excellent tracer of the photosphere of protoplanetary discs.
A famous example is the protoplanetary disc HD\,97048, whose flaring has been shown by   VLT Imager and Spectrometer for mid-InfraRed  (VISIR) imaging of the PAH emission at 8.6\,$\mu$m \citep{Lagage2006, Doucet2007}.\\
\\Although many protoplanetary discs show PAH features, there are several discs in which the IR PAH signatures are absent \citep{Meeus2001, vanBoekel2004, Acke2004}.
A study by \citet{Acke2010} shows that PAHs are detected in about 60--70\% of Herbig Ae/Be stars.
\citet{Valegard2021} show that 27\% of their sample discs around intermediate mass T\,Tauri stars (IMMT, defined as $1.5 M_\odot \leq M_* \leq 5 M_\odot$ and with a spectral type between F and K3) contain PAHs, where they adopted the criterion that at least two distinct PAH features must be present for a clear detection.
If only one feature is considered, the fraction of IMMT discs with PAHs increases to 44\%.
For classical T\,Tauri stars, \citet{Geers2006} find a lower detection limit of 8\%, excluding 14 tentative detections of PAHs in their total sample of 38 T\,Tauri discs.\\
\\Despite a very general correlation between PAH detection and stellar temperature due to the emitted UV field, it is not clear why the PAH emission strength also varies strongly within a stellar type.
To explain this phenomenon, it is necessary to look at protoplanetary discs individually and relate their morphology, environment, and properties to the PAH abundance and emission strength.
In the past several authors  proposed mechanisms for the variation of PAH strength in discs.
For the lack of PAH emission in T\,Tauri stars \citet{Siebenmorgen2010} and \citet{Siebenmorgen2012} modelled the destruction of PAHs through absorption of extreme ultraviolet (EUV) and X-ray photons. The authors show that for typical hard photon luminosities, PAHs can be sufficiently destroyed, and hence can explain the lack of PAH emission in these disc.
Other authors such as \citet{Geers2009} provide different possible explanations. Gas-phase PAHs can interact with icy dust grains and freeze out  on the grain surface.
When these dust grains are then exposed to UV radiation, the PAHs can be ionised, triggering ion-mediated reactions on the grain surface that deplete the PAHs \citep[e.g.][]{Ehrenfreund2006, Bouwman2010}.
Alternatively, in dense regions of the discs PAH molecules can aggregate to PAH clusters and form stack-like structures, as described by \citet{Rapacioli2005, Rapacioli2006}.
Due to their larger heat capacity, PAH clusters have a lower excitation temperature after absorption of a UV photon and consequently the strength of short-wavelength emission bands is reduced \citep{Bakes2001}.
Furthermore, the emission profile of PAH clusters is altered compared to the emission of individual PAH molecules since additional intermolecular modes occur in the MIR \citep{Rapacioli2009} that may be responsible for the broad underlying emission bands observed in the interstellar medium (ISM) \citep{Allamandola1989}. \\
\\In our previous study \citep{Lange2021} we  investigated the stability of PAH clusters that are fully exposed to the stellar UV radiation. We found that in the inner regions of the protoplanetary disc, PAH clusters dissociate within a very short time regardless of their size. 
In the outer regions of the disc, however,  clusters can no longer be photodissociated once they exceed a critical size as their peak temperature is too low to drive for dissociation.
As a result, we expect that PAH clustering in protoplanetary discs is a realistic scenario to surpress the emission of the short-wavelength PAH features, provided that the critical cluster size can be reached so that clusters can be stable in the disc photosphere.\\
\\In this study we want to investigate the  freeze-out of PAHs on dust grains and PAH clustering simultaneously as they are competing processes that  depend on the local gas-phase abundance of PAHs and grain density. 
We consider both processes as key processes to explain the absence of detectable PAH emission in some protoplanetary discs and focus our model on a Herbig Ae/Be star disc.
We want to quantify how the presence of dust grains in the protoplanetary disc affects the evolution and growth of PAH clusters and explore the possibility of using PAHs as an independent tool to obtain information on the dust grain population supplementing methods based on  submillimetre emission \citep[e.g. with ALMA][]{Birnstiel2018} and scattered light observations \citep[e.g.][]{Tazaki2019}.
In our models the gas-phase abundance of PAHs after its evolution in the disc is proportional to the total dust grain surface area. 
Therefore, if the PAH abundance of a protoplanetary disc is inferred from near-infrared (NIR) and MIR observations of the PAH features, our model is able to    check whether the dust population is consistent with ALMA/SPHERE characterisations of the dust grains.
Additionally, we model the desorption of PAH monomers and PAH clusters from the surface of  dust grains to investigate their ability to enter the gas-phase again.
As the James Webb Space telescope (JWST) was launched in December 2021, the NIR and MIR wavelength range will be more accessible than ever through the Mid-Infrared Instrument (MIRI) and the Near Infrared Camera (NIRcam) with high sensitivity and significant spatial resolution.
Hence, modelling work is crucial in order to understand the relevant processes and to analyse and interpret the upcoming observations.\\
\\This work is organised in the following way. In section \ref{sec:methods} we describe the coagulation and desorption models of the PAH clusters that were used.
In section \ref{sec:results} we present the results of our coagulation, desorption, and turbulent mixing model and investigate their implications on the PAH abundance in certain regions of the protoplanetary disc.
Then in section \ref{sec:discussion} we discuss the consistency of our results to previous studies, discuss our results in the context of sputtering, and heterogeneous PAH clusters.

%%%%%%%%%%%%%%%%%%%%%%%%%%%%%%%%%%%%%%%%%%%%%%%%%%%%%%%%%%%%%%%%%%%%%%%%%%%%%%
%%%%%%%%%%%%%%%%%%%%%%%%%%%%%%%%%%%%%%%%%%%%%%%%%%%%%%%%%%%%%%%%%%%%%%%%%%%%%%
%%%%%%%%%%%%%%%%%%%%%%%%%%%%%%%%%%%%%%%%%%%%%%%%%%%%%%%%%%%%%%%%%%%%%%%%%%%%%%
%%%%%%%%%%%%%%%%%%%%%%%%%%%%%%%%%%%%%%%%%%%%%%%%%%%%%%%%%%%%%%%%%%%%%%%%%%%%%%
%%%%%%%%%%%%%%%%%%%%%%%%%%%%%%%%%%%%%%%%%%%%%%%%%%%%%%%%%%%%%%%%%%%%%%%%%%%%%%

\section{Methods}
\label{sec:methods}

First, we     describe the relevant processes that influence the PAH abundance in our Herbig Ae/Be protoplanetary disc and describe where they take place in our model.
Here we neglect possible effects by hard photons (EUV, X-ray) as they are not dominant in Herbig star discs.
For this purpose, we divide the protoplanetary disc into layers depending on which process is important (figure \ref{fig:sketch}).
The upper layer of the disc is the photosphere where PAHs and PAH clusters are exposed to the full stellar radiation field and are able to desorb from dust grains. There, clusters are rapidly dissociated into their constituent monomers.
Below the photosphere, the UV field is partially shielded so that most clusters cannot be dissociated. 
However, it is strong enough so that dimers can be  dissociated more quickly  than they can grow from monomers inhibiting the growth of new clusters.
Hence, we call it the quiet layer as PAHs keep their current state.
Near the midplane, the coagulation layer is the region where PAHs rapidly coagulate into clusters and adsorb on dust grains.  We note that the coagulation layer includes not only   the classical cold midplane, but also the warmer layers above that are shielded sufficiently from UV radiation.\\
\\The coagulation layer is relatively dense and PAH clustering is very efficient.
Protected from stellar UV radiation, photodissociation of PAH clusters can be neglected there, and hence the coagulation layer provides an ideal place for the growth of PAH clusters.
Because dust particles are also present, PAH clusters will collide with them and freeze out to the dust particles.
As the adsorbed PAH is then coupled to the thermal reservoir of the dust particle, the clusters quickly adapt to the temperature of the grains and thermal evaporation of the clusters becomes impossible in most regions of the disc.
Therefore, we expect that in the coagulation layer PAHs are found exclusively as clusters on dust grains and that no PAHs are present in the gas phase.\\
\\In contrast, the photosphere is dominated by UV radiation from the central star.
Based on our previous study \citep{Lange2021}, the growth of PAHs to clusters is negligible as photodissociation is much faster than the first step of clustering, the formation of dimers through collisions of monomers.
Already existing clusters, however, can be dissociated and individual PAHs adsorbed on dust grains can be released from the grain surface when a UV photon is absorbed.
Further, adsorbed PAHs receive sufficient UV radiation such that desorption from the grain is possible.
As the photodissociation and photodesorption processes in the photosphere are strong, we do not expect cluster formation and freeze-out to be relevant there.\\
\\For each disc region we  developed a specific model taking into account the processes relevant there.
To understand the clustering and adsorption of PAHs on dust grains in the coagulation layer, we  developed a coagulation code, which we describe in section \ref{sec:coag}.
We then present our model for the desorption of clusters and monomers in the photosphere (section \ref{sec:desorp}), which we later apply to the results of the coagulation code to determine what fraction of the original PAHs can be recovered from the dust grains when they are mixed from the coagulation layer into the photosphere.

%%%%%%%%%%%%%%%%%%%%%%%%%%%%%%%%%%%%%%%%%%%%%%%%%%%%%%%%%%%%%%%%%%%%%%%%%%%%%%
%%%%%%%%%%%%%%%%%%%%%%%%%%%%%%%%%%%%%%%%%%%%%%%%%%%%%%%%%%%%%%%%%%%%%%%%%%%%%%
%%%%%%%%%%%%%%%%%%%%%%%%%%%%%%%%%%%%%%%%%%%%%%%%%%%%%%%%%%%%%%%%%%%%%%%%%%%%%%

\subsection{PAH coagulation model}
\label{sec:coag}
The growth of PAHs into clusters is a mechanism that reduces the feature-to-continuum ratio of PAH bands \citep{Geers2009} because temperature fluctuations are reduced compared to individual monomers.
Furthermore, the additional vibrational modes of PAH clusters \citep{Rapacioli2007} are responsible for underlying broad bands below the PAH features \citep{Allamandola1989} which further reduce the observable feature-to-continuum ratio.
So far, we have only investigated photodissociation of clusters  \citep{Lange2021} and estimated the cluster growth timescale based on isolated gas-phase PAH collision rates.
In this work we want to determine how fast PAH clusters can grow in the coagulation layer given that dust grains are also able to interact with PAHs through collisions.
On the one hand, the deposition of PAHs onto dust grains is another mechanism for reducing the strength of the MIR PAH bands, as the adsorbed PAHs are thermally coupled to the dust grains, and therefore have lower peak temperatures.
On the other hand, the presence of dust grains allows PAHs to stick to the surface of the grains so that these PAHs are no longer available for PAH cluster formation in the gas-phase.
Since the coagulation layer is shielded from UV radiation, we neglect photodissociation of forming PAH clusters in this model. 
Furthermore, the expected temperature in the coagulation layer is lower than the required 400\,K (for the smallest simulated PAH coronene) above which thermal desorption and thermal dissociation become relevant, and therefore we also neglect these processes.
In our coagulation model we consider a local box filled with PAHs, dust, and gas without transport of PAHs and dust in the  radial or vertical direction because the coagulation is much faster than the transport processes, as we  show below (see figure \ref{fig:PAHgrowth}).\\
\\We describe the size of PAH clusters by the number of carbon \textit{atoms} $N_\text{C}$ (not monomer molecules) it contains.
Then the cluster growth can be described by the classical Smoluchowski equation \citep{Smoluchwoski1916}
\begin{equation}
\begin{split}
    \frac{df(N_\text{C})}{dt} &= \int_0^{N_\text{C}/2} f(N_\text{C}')f(N_\text{C}-N_\text{C}')k(N_\text{C}',N_\text{C}-N_\text{C}')dN_\text{C}'\\
     &- \int_0^\infty f(N_\text{C}) f(N_\text{C}') k(N_\text{C},N_\text{C}')dN_\text{C}'\\
     &- Q_-(N_\text{C}),
     \end{split}
\label{eq:coag}
\end{equation}
where $f(N_\text{C},r,\theta)$ is the number density of a given PAH cluster size at a given radial position $r$ and height $\theta$;
$k$ describes the coagulation kernel; and $Q_-$ is a sink term to account for the adsorption of PAHs from the gas phase to grains.
We do not consider the case that PAHs can grow into clusters on grains here.  
The first term in eq. \eqref{eq:coag} describes the formation of a cluster with $N_\text{C}$ carbon atoms from the collision of two smaller clusters; the second term describes the loss of clusters with size $N_\text{C}$ through collisions with other clusters.
At van der Waals binding energies of 1 eV between PAH monomers and the cluster, very high temperatures of more than 13\,500\,K are needed for destructive monomer-cluster collision outcomes \citep{Rapacioli2006}.
With the relevant thermal energies in the protoplanetary disc, we can safely assume perfect sticking for PAHs monomer-monomer, monomer-cluster, cluster-cluster, and all PAH-grain collisions.\\
\\For PAHs in the gas phase, we only consider Brownian motion and neglect turbulent processes for the coagulation model as sources for relative velocities as these do not contribute, due to the strong coupling of PAHs to the gas.
The coagulation kernel can then be determined by the relative velocity $\Delta v_{i,j}$ and the collision cross-section $\sigma_{i,j}$, which both depend on the size of the collision partners $i$ and $j$ involved:
\begin{equation}
    k = \sigma_{i,j} \Delta v_{i,j} \text{.}
    \label{eq:kernel}
\end{equation}
The relative velocities in Brownian motion are given by 
\begin{equation}
    \Delta v_{i,j} = \sqrt{\frac{8k_\text{B}T}{\pi\mu_{i,j}}}
    \label{eq:deltav}
,\end{equation}
with $\mu_{i,j}$ being the reduced mass of the collision partners,  $k_\text{B}$ the Boltzmann
constant, and $T$   the kinetic temperature of the gas that we set to $T=100$\,K.
The collision  cross-section $\sigma_{i,j}$ between PAH clusters is assumed to be the maximum geometrical cross-section with radius $r$
\begin{equation}
    \sigma_{i,j} = \pi (r_i + r_j)^2 \text{.}
\end{equation}

%%%%%%%%%%%%%%%%%%%%%%%%%%%%%%%%%%%%%%%%%%%%%%%%%%%%%%%%%%%%%%%%%%%%%%

\subsubsection{PAH cluster sizes and adsorption}
\label{sec:cluster_size}
Based on the studies of \citet{Rapacioli2005}, small PAH clusters tend to prefer stack-like structures, while larger clusters favour extended 3D structures made from smaller stacks.
However, since the temperature of a cluster is never zero, other configurations are energetically possible, and even small clusters can form modified stack structures or 3D structures \citep{Rapacioli2007}.
Consideration of all the different allowed cluster structures would unnecessarily complicates the model.
Hence, we decided to simplify the model and divide the effective radius into three size ranges: monomers, which are a planar molecules;
small clusters, which grow linearly to account for the formation of stacks until the diameter of a monomer is equal to the length of a stack; and large clusters, which grow spherically as compact dust grains.
The detailed formulas for the cluster radius $r_i$ over the entire size range can be found in Appendix \ref{app:cluster_size}.\\
\\As we did for the growth of PAH clusters, we model adsorption on dust grains with equation \eqref{eq:kernel} and \eqref{eq:deltav}, where the collision partners here are a PAH cluster $i$ and a dust grain $l$ with grain size $a_l$.
We take this into account in a semi-analytical approach by summing over all grain sizes $l$,
\begin{equation}
    Q_-(N_\text{C}) = f(N_\text{C})\int_{a_\text{min}}^{a_\text{max}} k\,n(a)\,da
    \label{eq:Q-}
,\end{equation}
where $n_l$ is the number density of the dust grains.
Consequently, the adsorption of PAHs depends on both the local dust grain abundance and the grain size distribution.
Therefore, to characterise the dust population with a single parameter, we  use the total collisional cross-section per unit volume of the dust as the relevant quantity defined by
\begin{equation}
    \sigma_\text{dust} = \int_{a_\text{min}}^{a_\text{max}} \pi n(a) a^2 da
.\end{equation}
We note that this approach is limited to the case where the PAH clusters are much smaller than the dust grains ($r_i \ll a$), which is unrestrictedly valid in our set-up.\\
\\To  solve the coagulation equation \eqref{eq:coag}, we follow the principles described in the Appendix of \cite{Dullemond2005}. 
To ensure mass conservation, we use their conservation method, but apply it only to the two integral terms where numerical mass loss may inadvertently occur.
We add for the time step determination a safety factor of 0.3 and stop the simulation as soon as less than $10^{-4}$\,\% of the PAH mass is left in the gas phase.
We start our simulation with PAH monomers in the gas phase only, where $N_\text{C,0}$ describes the number of carbon atoms in a monomer. To ensure numerical stability, we use a Gaussian of width $\sigma=5$\,C atoms centred around $N_\text{C,0}$ as our initial distribution.

%%%%%%%%%%%%%%%%%%%%%%%%%%%%%%%%%%%%%%%%%%%%%%%%%%%%%%%%%%%%%%%%%%%%%%%%%%%%%%
%%%%%%%%%%%%%%%%%%%%%%%%%%%%%%%%%%%%%%%%%%%%%%%%%%%%%%%%%%%%%%%%%%%%%%%%%%%%%%
%%%%%%%%%%%%%%%%%%%%%%%%%%%%%%%%%%%%%%%%%%%%%%%%%%%%%%%%%%%%%%%%%%%%%%%%%%%%%%

\subsection{Photodesorption of PAH clusters from dust grains}
\label{sec:desorp}
After presenting the coagulation model, we   look at PAHs frozen onto dust grains and how they can be brought back into the gas phase.
Given that cluster formation and freeze-out require a strong UV shielding in the coagulation layer to be faster than their counter processes, desorption must occur higher up in the disc.
Additionally, PAH charging through photo-ionisation and pick-up of free-electrons (i.e. from C->C+) lead to a difficult PAH charge distribution \citep{Thi2019} that can prevent PAH sticking higher up in the disc. 
Assuming a turbulent disc, the PAH-bearing dust grains need to be transported vertically through the disc by turbulence until they eventually reach the UV-rich photosphere to desorb from the grains.
When the equilibrium temperature of the grains is above the evaporation temperature of the PAHs, PAHs can thermally evaporate.
In addition, we consider that adsorbed PAH clusters absorb UV photons that increase the internal excitation of the PAH vibrational modes, including those that bind the PAHs to the grain.
This internal excitation will be lost through coupling to the grain phonon modes.
Therefore, the competition between sublimation and energy transfer to the grain must be studied to determine the sublimation rate.
Due to the high heat capacity of the dust grains, the temperature fluctuations after absorption of a photon by the dust grains are too small to sublimate PAHs. 
Only direct photon absorptions by the PAHs can lead to sublimation.\\
\\Like the dissociation of PAH clusters by UV radiation, the sublimation of PAHs can be simulated by a Monte Carlo model where we consider photon absorption, cooling, and desorption of the entire cluster as possible events.

%%%%%%%%%%%%%%%%%%%%%%%%%%%%%%%%%%%%%%%%%%%%%%%%%%%%%%%%%%%%%%%%%%%%%%%%%%%%%%

\subsubsection{Photon absorption}
We model the absorbed energy flux of a PAH $\Phi_{\text{a},\text{E}}$ at a given photon energy $E$ by
\begin{equation}
    \Phi_{\text{a},\text{E}} = F_\text{E} \sigma_\text{E}
,\end{equation}
where $\sigma_\text{E}$ is the photon absorption cross-section of a PAH at photon energy $E$, and $F_\text{E}$ is the stellar flux density.
We estimate the stellar flux in the photosphere with a black-body spectrum with stellar temperature and luminosity, as given in Table \ref{tab:disc_models}.
To determine the flux deeper in the disc, we moderate the flux with an optical depth through the radiative transfer code RADMC-3D \footnote{\url{https://www.ita.uni-heidelberg.de/~dullemond/software/radmc-3d/contributions.php}} and the opacity calculator OpTool \citep{Dominik2021} using the standard DIANA opacities \citep{Woitke2016}.
For $\sigma_\text{E}$ we use photon absorption cross-sections from time dependent density functional theory (DFT) calculated by \citet{Malloci2007} for individual PAH species ranging from coronene (C$_{24}$H$_{12}$) to circumovalene (C$_{66}$H$_{20}$).
We divide the energy space between 0.1-13.6\,eV into 100 bins with average energy $E_m$ so that we can derive an absorption rate of
\begin{equation}
    r_{\text{a}, m} = \frac{1}{E_m} \int_{E_\text{i, min}}^{E_\text{i,max}} \Phi_{\text{a}, \text{E}} \text{d}E \text{.}
\end{equation}
We do not consider photon energies above 13.6\,eV. Our assumed black-body spectrum for a typical Herbig star does not include a significant amount of EUV or X-ray photons with energies of a few 100\,eV or more.
Therefore, the ionisation of hydrogen increases the optical depth around 13.6\,eV significantly so that the relevant photons with slightly higher energies cannot penetrate deeper into the disc \citep{Ryter1996}, and therefore do not contribute to the evolution of PAHs.

%%%%%%%%%%%%%%%%%%%%%%%%%%%%%%%%%%%%%%%%%%%%%%%%%%%%%%%%%%%%%%%%%%%%%%%%%%%%%%

\subsubsection{Cooling through thermal coupling to grains}
Cooling of PAH clusters can happen by IR photon emission and by energy transfer to the dust grain.
However, since cooling by IR photons happens on a timescale of seconds \citep{Bakes2001} and is thus much slower than energy transfer to the dust grain, we can neglect IR cooling.
We further assume that the temperature of the dust grains is constant at the radiative equilibrium value because the grains are large enough to neglect temperature fluctuations \citep{Tielens2005}. 
Since we focus on the optically thin photosphere, we use equation \eqref{eq:T_dust} to estimate the equilibrium temperature of the grains.\\
\\For the van der Waals bond between the grain and the PAHs (resp. PAH clusters), we assume that there is a flat contact surface between the grain-facing PAH and the dust grain so that the potential energy is minimised. Then we can assume that the energy transfer can be modelled as heat transfer between two graphene sheets through the c-plane.
We expect the PAH and grain temperature to be in the range of 100-1000\,K, and hence we assume a thermal conductivity of $\kappa = 0.04$\,W/m/K = 4\,000 erg/s/cm/K based on the theoretical calculations for graphene by \citet{Alofi2014}.
The effective contact surface is determined by the size of the PAH monomer species through
\begin{equation}
    A = \pi \left(\sqrt{N_\text{C,0}} \cdot 0.9 \text{\AA}\right)^2 = N_\text{C,0} \cdot 2.5 \cdot 10^{-16} \text{\,cm}^2
,\end{equation}
where $N_\text{C,0}$ is the number of C atoms of a PAH monomer \citet{Tielens2008}.
The energy transfer rate from the PAH cluster to the grain is then determined as 
\begin{equation}
    \frac{\text{d}Q}{\text{d}t} = \frac{\kappa A \Delta T}{\Delta d}
    \label{eq:conduction}
,\end{equation}
where $\Delta d$ is the bond length between the surface molecule and the grain surface.
We assume the typical alternating graphite interlayer distance of $\Delta d = 3.34\,\AA$ \citep{Andres2008} for the bond length, which is comparable to the van der Waals bond distance of two plane-parallel PAH molecules of $\approx 3.5\,\AA$  \citep{Rapacioli2005}.
The energy transport from the cluster to the dust grain is not a classical macroscopic heat transport.
Instead, energy is transferred quantum-wise with the natural frequency of the van der Waals mode of the bond that can be estimated by 
\begin{equation}
    v_z = \sqrt{\frac{2 N_\text{s} E_\text{A}}{\pi^2 m}}
,\end{equation}
where $N_\text{s} \approx 2 \cdot 10^{15}$ sites/cm$^2$ is the number of binding sites of the adsorbent, $E_\text{A}$ is the binding energy, and $m$ the mass of the adsorbate \citep{Tielens2005}.
For a coronene monomer with a binding energy of $E_\text{A} = 1.4$\,eV, the corresponding frequency is $v_z = 1.4\cdot 10^{12}\,\text{s}^{-1}$.
To match the vibrational frequency, we choose the energy of a heat quantum to be 1\,\% of the remaining excitation energy of the cluster ($\Delta E = 0.01 (E-E_\text{grain})$) so that the cooling rate
\begin{equation}
    r_\text{c} = \frac{\text{d}Q}{\text{d}t}\Delta E^{-1} \approx 10^{12}\,\text{s}^{-1}
\end{equation}
agrees with the estimated van der Waals frequency.
We stop the cooling process as soon as the temperature of the cluster deviates less than 0.1\,K from that of the grain and use the equilibrium temperature of the dust grain until the next absorption of a photon to improve the efficiency of the code.

%%%%%%%%%%%%%%%%%%%%%%%%%%%%%%%%%%%%%%%%%%%%%%%%%%%%%%%%%%%%%%%%%%%%%%%%%%%%%%

\subsubsection{Evaporation rate}
The evaporation of adsorbates can be generally described by the Arrhenius equation, which can be derived from the Rice-Ramsperger-Kassel-Marcus (RRKM) theory \citep{Tielens2005}.
The classical expression is
\begin{equation}
    r_\text{k} = k_0\cdot \text{exp}\left(\frac{-E_\text{A}}{k_\text{B}T_\text{e}}\right)
    \label{eq:arrhenius}
,\end{equation}
where $k_0$ is the evaporation rate constant, $E_\text{A}$ the binding energy of the PAH cluster to the grain, and $T_\text{e}$ is the effective temperature of the PAH cluster determined by
\begin{equation}
    T_\text{e} = T_\text{m} \left(1-0.2\frac{E_\text{A}}{E}\right) \text{\,.}
    \label{eq:Teff}
\end{equation}
We assume $k_0 = 2.5\cdot10^{17}$\,s$^{-1}$ for all PAH clusters similar to the dissociation of PAH clusters derived in \citet{Lange2021}.
We use the heat capacity of \citet{Bakes2001}, and thereby describe the temperature-energy relation as \citep{Tielens2021}
\begin{equation}
    T_\text{m} = \begin{cases}
    3750 \left(\frac{E\text{(eV)}}{3N-6}\right)^{0.45}\text{\,K}\text{\hspace{1cm} if $T_m < 1000$\,K}\\
    11000 \left(\frac{E\text{(eV)}}{3N-6}\right)^{0.8}\text{\,K}\text{\hspace{1cm} if $1000\text{\,K} < T_m$}\\
    \end{cases}
    \label{eq:T(E)}
,\end{equation}
where $N$ is the total number of atoms in the cluster.
We estimate the activation energy $E_A$ for the evaporation of PAH clusters from dust grains by using an approximation from DFT calculations for the adsorption of PAHs on a graphene surface by \citet{Li2018} 
\begin{equation}
    \frac{E_\text{A}}{N_\text{C}} = 0.046+0.021\frac{N_\text{H,0}}{N_\text{C,0}}
\label{eqEA}
,\end{equation}
with $N_\text{H,0}$ and $N_\text{C,0}$ being respectively the number of hydrogen and carbon atoms in a PAH monomer. This approximation matches the findings of the experimental study by \citet{Zacharia2004}.
The calculated binding energies can be found in Table \ref{tab:PAH_props}.
Our derived evaporation rates agree with \citet{Montillaud2014}, who interpolated evaporation rates for clusters calibrated through molecular dynamics studies.

%%%%%%%%%%%%%%%%%%%%%%%%%%%%%%%%%%%%%%%%%%%%%%%%%%%%%%%%%%%%%%%%%%%%%%%%%%%%%%

\subsubsection{Monte Carlo scheme}
Finally, the desorption rate is determined by evaluating the energy fluctuations of the PAH clusters after absorption of UV photons with the desorption rates.
The energy probability distribution $G(E)$ of the adsorbed PAH cluster is determined by a Monte Carlo simulation in which photon absorption, heat transport, and desorption events are considered.
We do this using a Monte Carlo scheme that follows the fundamental principles explained in \citet{Zsom2008}.
The total rate of events for a simulated PAH molecule is thus
\begin{equation}
    r = \sum_m r_{\text{a}, m} + r_\text{c} + r_\text{k} \text{\,.}
\end{equation}
We note that although multi-photon absorption events are included in this equation, they are extremely rare because the energy transfer from the PAHs to the grain ($\tau \approx 10^{-10}$\,s) is very fast compared to the absorption rate ($ \tau \approx 1$\,s) of the UV photons. 
The energy probability distribution is therefore entirely dominated by single photon absorptions.
One Monte Carlo time step $\delta t$ is determined by a random draw from an exponential distribution with mean $1/r$
\begin{equation}
    f(\delta t) = r \,\text{exp}(-r\delta t) \text{.}
\end{equation}
We determine which event has occurred by the set of all possible events with rates $R = \{ r_\text{{a},m}$, $r_\text{c}, r_\text{k}\}$
with the probability of an event $j$ with rate $r_j \in R$ given by
\begin{equation}
    P(j) = \frac{r_j}{r} \text{.}
\end{equation}
The final effective desorption rate can be determined by integrating all possible energy states of $G(E)$ with the individual desorption rates $r_\text{k}$:
\begin{equation}
    k = \int_0^{\infty} G(E)r_\text{k}(E) \text{d}E.
    \label{eq:k_int}
\end{equation}

%%%%%%%%%%%%%%%%%%%%%%%%%%%%%%%%%%%%%%%%%%%%%%%%%%%%%%%%%%%%%%%%%%%%%%%%%%%%%%
%%%%%%%%%%%%%%%%%%%%%%%%%%%%%%%%%%%%%%%%%%%%%%%%%%%%%%%%%%%%%%%%%%%%%%%%%%%%%%
%%%%%%%%%%%%%%%%%%%%%%%%%%%%%%%%%%%%%%%%%%%%%%%%%%%%%%%%%%%%%%%%%%%%%%%%%%%%%%

\subsection{The disc model}
\begin{table}[]
    \centering
    \caption{Properties of the standard Herbig Ae/Be disc model (STD) used for the calculations and alternative dust grain distributions. In all models the following properties are used: inner disc radius $r_\text{in}=0.5$\,au, outer disc radius $r_\text{out}=100$\,au, disc mass $m_\text{disc}=0.228 M_\odot$, stellar mass $M_*=2.28 M_\odot$, stellar luminosity $L_*=15.33 L_\odot$, and stellar temperature $T_*=8200$\,K. The stellar properties (reference star HD\,169142; \citealt{Honda2012}) are chosen to be comparable to the photodissociation results from our previous study in \citet{Lange2021}. For comparison of other grain populations, we use models to account for only small grains (SGs), only large grains (LGs), and an extreme case with only very large grains (VLGs), where $\sigma_\text{PAH}$/$\sigma_\text{dust}$ is given in the midplane at 10\,au. We find no significant deviation of $\sigma_\text{PAH}$/$\sigma_\text{dust}$ within the disc at relevant heights for coagulation similar to fig. \ref{fig:sigma-sigma}. for the alternative grain distributions.}
    \begin{tabular}{c|c|c|c|c}
         Name & $a_\text{min}$& $a_\text{max}$ & $\gamma$ & $\sigma_\text{PAH}$/$\sigma_\text{dust}$\\ \hline
         STD & 0.05\,$\mu$m & 1\,mm & 3.5 & 310\\
         SG & 0.05\,$\mu$m & 1\,$\mu$m & 3.5 & 1.7\\
         LG & 1\,$\mu$m & 1\,mm & 3.5 & 6300\\
         VLG & 10\,$\mu$m & 10\,mm & 3.5 & 14000\\
    \end{tabular}
    \label{tab:disc_models}
\end{table}
The values of dust density and dust size distribution in the protoplanetary disc are crucial in our coagulation model, while temperature and radiation intensity are essential for the evaporation of PAHs from dust grain surfaces.
Therefore, we need to set up a disc model to define all relevant quantities.
Our disc has a gas surface density profile of
\begin{equation}
    \Sigma(r) = 730 \frac{\text{g}}{\text{cm$^2$}} \left(\frac{r}{1\,\text{au}}\right)^{-1.5} \left(\frac{M_*}{M_\odot}\right)
\end{equation}
scaled to a minimum mass solar nebula (MMSN) power-law profile, as first described by \citet{Weidenschilling1977}.
Our disc is in vertical hydrostatic equilibrium,
\begin{equation}
    \rho(z,r) = \rho_\text{mid}(r)\,\text{exp}\left(\frac{-z^2}{2H^2}\right)
,\end{equation}
where the gas pressure scale height is determined by $H = c_\text{s}/\Omega_\text{K}$.
The sound speed $c_\text{s}$ is calculated by $c_\text{s}=\sqrt{k_\text{B}T/\mu m_\text{p}}$, where we assumed $\mu = 1.37$ and an optically thin surface layer so that the temperature profile can be described through the approach of \citet{Hayashi1981}:
\begin{equation}
    T(r) = 280\,\text{K} \left(\frac{r}{\text{au}}\right)^{-0.5}\left(\frac{L_*}{L_\odot}\right)^{0.25} \text{.}
    \label{eq:T_dust}
\end{equation}
The midplane density is determined by $\rho_\text{mid}(r) = \Sigma(r)/\sqrt{2\pi}H(r)$.\\
\\For the dust grains, we assume a Mathis-Rumpl-Nordsieck (MRN) size distribution \citep{Mathis1977}. Then the number density of the dust grains of size $a$ is given by
\begin{equation}
    n(a) \propto a^{-\gamma}
    \label{eq:dust_distro}
,\end{equation}
with $\gamma = 3.5$.
Consequently, the smallest dust grains provide the largest share of the available surface area, and are therefore the dominant Brownian-motion collision partners for PAH molecules. Hence, PAHs are mainly adsorbed on the smallest available dust grains.
In our standard model these grains are also closely coupled to the gas.
However, the large dust grains also contribute to the total dust surface area.
We want to consider an evolved dust disc that is in equilibrium between settling and mixing.
To do this we use the results of \citet{Fromang2007} who made magnetohydrodynamic (MHD) simulations and apply their fitted turbulent toy model as an estimate for our disc.
In their model the dust grains do not exert any feedback on the gas, which is why we use the equations for the hydrostatic equilibrium for the gas density.
The vertical density distribution of the dust grains can be described by
\begin{equation}
    \frac{\partial\rho_\text{d}}{\partial t} - \frac{\partial}{\partial z}\left(z \Omega^2 \tau_\text{s}\rho_\text{d}\right) = \frac{\partial}{\partial z} \left[D\rho \frac{\partial}{\partial z}\left(\frac{\rho_\text{d}}{\rho}\right)\right]
,\end{equation}
where the steady-state solution is given through
\begin{equation}
    \frac{\partial}{\partial z}\left(\text{ln}\frac{\rho_\text{d}}{\rho}\right) = - \frac{\Omega^2 \tau_\text{s}}{D} z \text{.}
    \label{eq:dust}
\end{equation}
The diffusion coefficient $D$ of the dust grains is defined by the local velocity fluctuations
$\delta v_\text{z}$
\begin{equation}
    D = \delta v_\text{z}^2\tau_\text{corr}
,\end{equation}
where the correlation time of the velocity fluctuations $\tau_\text{corr}$ is determined by 
\begin{equation}
    \tau_\text{corr} = 0.15\frac{2\pi}{\Omega_\text{k}} \text{.}
\end{equation}
From their hydrodynamical studies, \citet{Fromang2007} found that $\delta v_\text{z}$ can be best described by
\begin{equation}
    \delta v_\text{z} = 
    \begin{cases}
    \delta v_\text{z, mid} + [\delta v_\text{z,up} - \delta v_\text{z,mid}] \left(\frac{|z|}{2H}\right)^2 &\text{    if } |z| < 2H\\
    \delta v_\text{z, up} &\text{    else}
    \end{cases}
,\end{equation}
with $\delta v_\text{z,up} = 0.15 c_\text{s}$ and $\delta v_\text{z, mid} = 0.05 c_\text{s}$.
Finally, the stopping time of the particles $\tau_\text{s}$ is defined as 
\begin{equation}
    \tau_\text{s} = \frac{\rho_{s}a}{\rho c_\text{s}}\text{.}
    \label{eq:tau}
\end{equation}
We numerically solve equation \eqref{eq:dust} to obtain the local dust grain density $\rho_\text{d}(a,r,\theta)$. To ensure numerical stability we solve for $\text{ln}(\rho_\text{d}/\rho)$ rather than $\rho_\text{d}$. The parameters used for the disc set-up are provided in Table \ref{tab:disc_models}. Additionally, a test case set-up to compare to \citet{Fromang2007} is provided in Appendix \ref{app:dust_density}.

\begin{figure}
    \centering
    \includegraphics[width=0.99\linewidth]{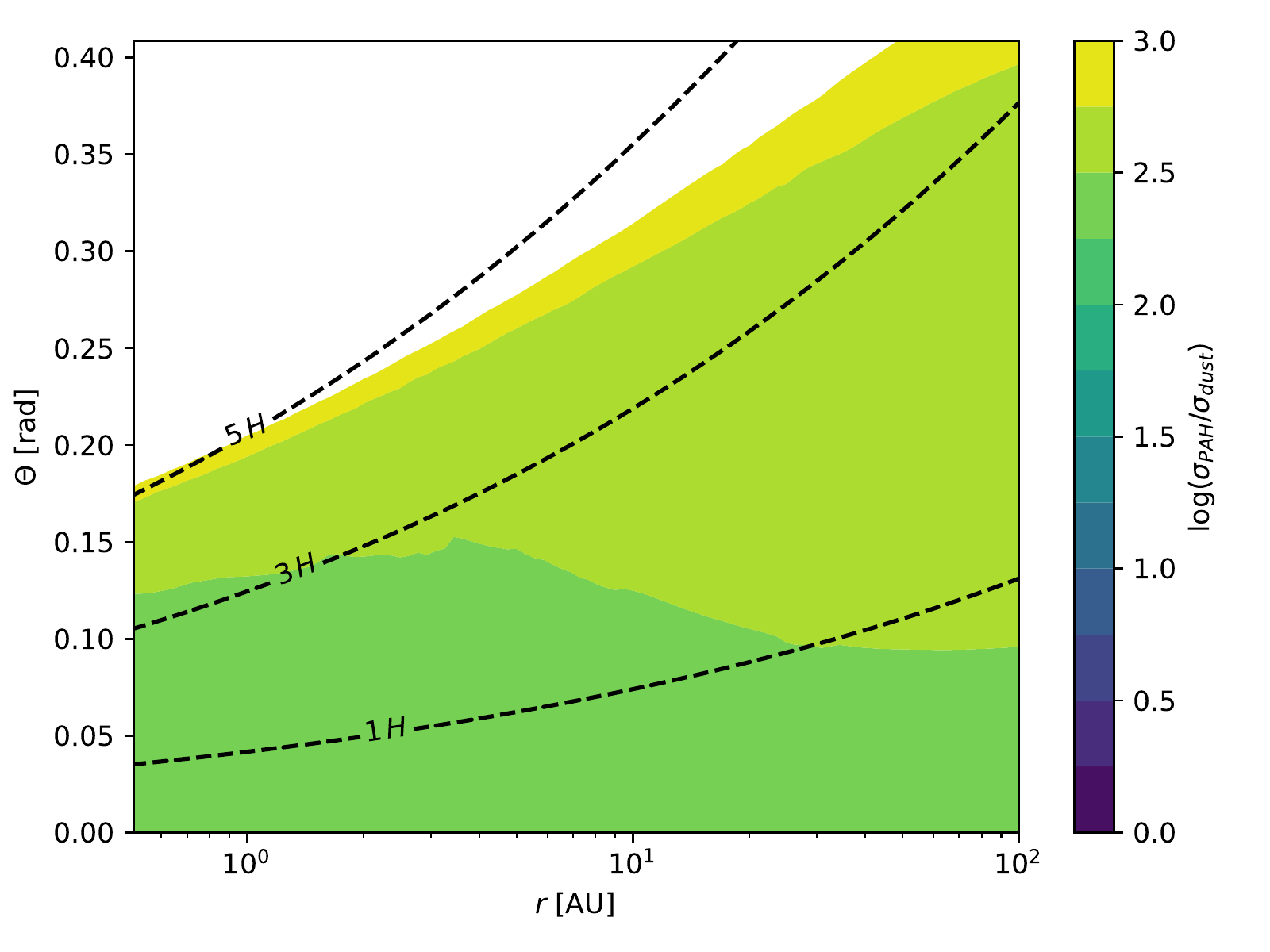}
    \caption{Ratio of total PAH collisional cross-section $\sigma_\text{PAH}$   to the total dust collisional cross-section $\sigma_\text{dust}$ for the standard Herbig disc model STD with ISM PAH abundance. As the surface area of the dust is dominated by the smallest grains, which are closely coupled to the gas, the ratio is almost constant $\sigma_\text{PAH}/\sigma_\text{dust}\approx 310$ resp. log$\left(\sigma_\text{PAH}/\sigma_\text{dust}\right)\approx 2.5$ at relevant coagulation altitudes ($z < 3H$). Only in the outer disc does this ratio change; however, the relevant coagulation height decreases as well (see figure \ref{fig:layererd_disk}). The gas pressure scale heights of $z$ = 1\,$H$, 3\,$H$, 5\,$H$ are indicated by dashed lines.}
    \label{fig:sigma-sigma}
\end{figure}

\subsubsection{Vertical mixing}
\label{sec:turbulence}
In order to estimate how likely the evaporation of PAH clusters is in a turbulent disc, we want to compare the evaporation rate with the mean-residence time $t_\text{mrt}$ of an adsorbed PAH cluster in the photosphere given a typical $\alpha$-viscosity disc \citep{Shakura1973}.
For this purpose we use the standard turbulence model typically used in the models of \citet{Dubrulle1995},  \citet{Dullemond2004}, \citet{Schraepler2004}, and \citet{Fromang2007},  among others.
In this model, the turbulent velocities of the largest vortices are proportional to the speed of sound $c_\text{s}$. Then the turbulent turnover time can be estimated by
\begin{equation}
    t_\text{ed} = \frac{\alpha^{1-2q}}{\Omega_\text{k}}
    \label{eq:t_turn}
,\end{equation}
with the preferred scaling of $q = 0.5$ so that the length scale of an eddy is given by
\begin{equation}
    l_\text{ed} = \alpha^{0.5} H \text{ .}
\end{equation}
We estimate that the probability of a grain coupling to a neighbouring eddy is given by the mass fluxes through the foot $j_z$ and top $j_{z+l_\text{ed}}$ of the eddy  via
\begin{equation}
    j(z) = v \rho(z) = \alpha^{0.5} c_\text{s} \rho(z)
\end{equation}
evaluated at the foot and top of an eddy.
The probabilities can then be evaluated through mass conservation respectively by
\begin{equation}
    p_\text{up}(z) = \frac{j(z+l_\text{ed})} {j(z) + j(z+l_\text{ed})}
    \label{eq:j_prob}
\end{equation}
and
\begin{equation}
    p_\text{down}(z) = \frac{j(z)} {j(z) + j(z+l_\text{ed})} \text{.}
    \label{eq:j_prob2}
\end{equation}
Using the $\alpha$ description model for the mass flux, the mixing probabilities become vertically  dependent on the densities at the given height so that equations \eqref{eq:j_prob} and \eqref{eq:j_prob2} respectively become
\begin{equation}
    p_\text{up}(z) = \frac{\rho(z+l_\text{ed})} {\rho(z) + \rho(z+l_\text{ed})}
\end{equation}
and 
\begin{equation}
p_\text{down}(z) = \frac{\rho(z)} {\rho(z) + \rho(z+l_\text{ed})}
    \label{eq:rho_prob}
,\end{equation}
which under hydrostatic equlibrium reduce to a simple exponential form.
By tracing the particle movement through the eddies, we can determine how long carried particles are fully exposed to the incident UV radiation in the photosphere and how often they mix between the photosphere and the disc midplane. We follow a particle through $10^8$ randomly sampled steps in the disc with an upper boundary of $4 H_p$ where particles are forced to mix down.
%%%%%%%%%%%%%%%%%%%%%%%%%%%%%%%%%%%%%%%%%%%%%%%%%%%%%%%%%%%%%%%%%%%%%%%%%%%%%%
%%%%%%%%%%%%%%%%%%%%%%%%%%%%%%%%%%%%%%%%%%%%%%%%%%%%%%%%%%%%%%%%%%%%%%%%%%%%%%
%%%%%%%%%%%%%%%%%%%%%%%%%%%%%%%%%%%%%%%%%%%%%%%%%%%%%%%%%%%%%%%%%%%%%%%%%%%%%%
%%%%%%%%%%%%%%%%%%%%%%%%%%%%%%%%%%%%%%%%%%%%%%%%%%%%%%%%%%%%%%%%%%%%%%%%%%%%%%
%%%%%%%%%%%%%%%%%%%%%%%%%%%%%%%%%%%%%%%%%%%%%%%%%%%%%%%%%%%%%%%%%%%%%%%%%%%%%%

\section{Results}
\label{sec:results}
\subsection{Coagulation and freeze-out of PAH clusters}
\label{sec:res_coag}
\begin{figure}
    \centering
    \includegraphics[width=0.99\linewidth]{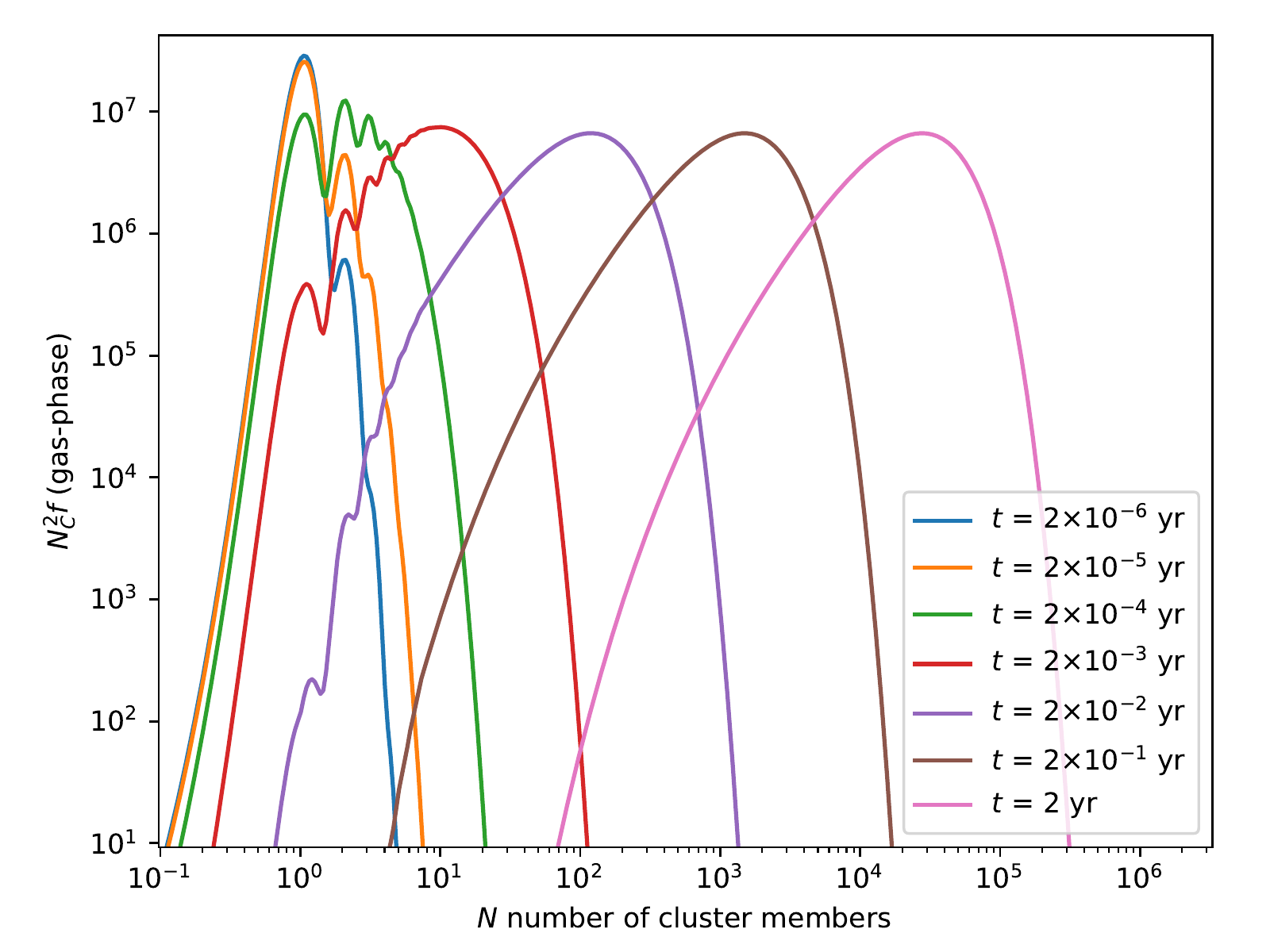}
    \caption{PAH coagulation model without the presence of dust grains at $r=10$\,au and $\theta=0$ in the coagulation layer for an initial ISM PAH abundance of coronene monomers. The  distribution is shown for selected time cuts after the start of the simulation. The growth of PAH monomers to very small carbonaceous grains is very fast, and within years only the largest clusters remain.}
    \label{fig:PAHgrowth}
\end{figure}

\begin{figure}
    \centering
    \includegraphics[width=0.99\linewidth]{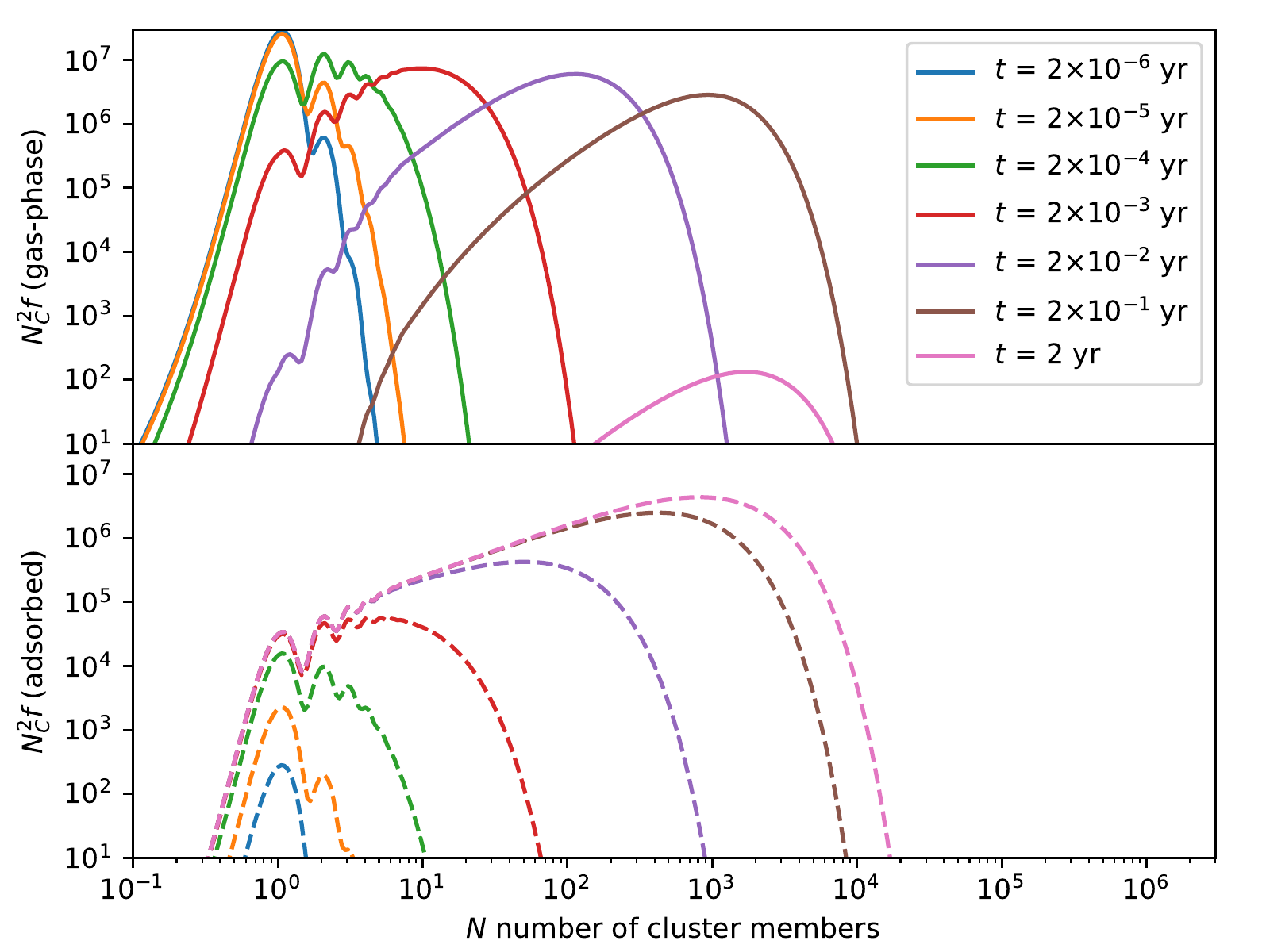}
    \caption{Similar to figure \ref{fig:PAHgrowth} with the same time cuts, but in the presence of dust grains. The dashed lines show the size distribution of adsorbed PAH clusters. Initially, the density of PAHs is high enough that clustering is faster than adsorption. Over time, PAHs get adsorbed on dust grains until adsorption and clustering act on similar timescales. Then growth is halted, and clusters will be stuck on grains. Most PAHs are adsorbed as large clusters.}
    \label{fig:PAHgrowthDust}
\end{figure}

\begin{figure}
    \centering
    \includegraphics[width=0.99\linewidth]{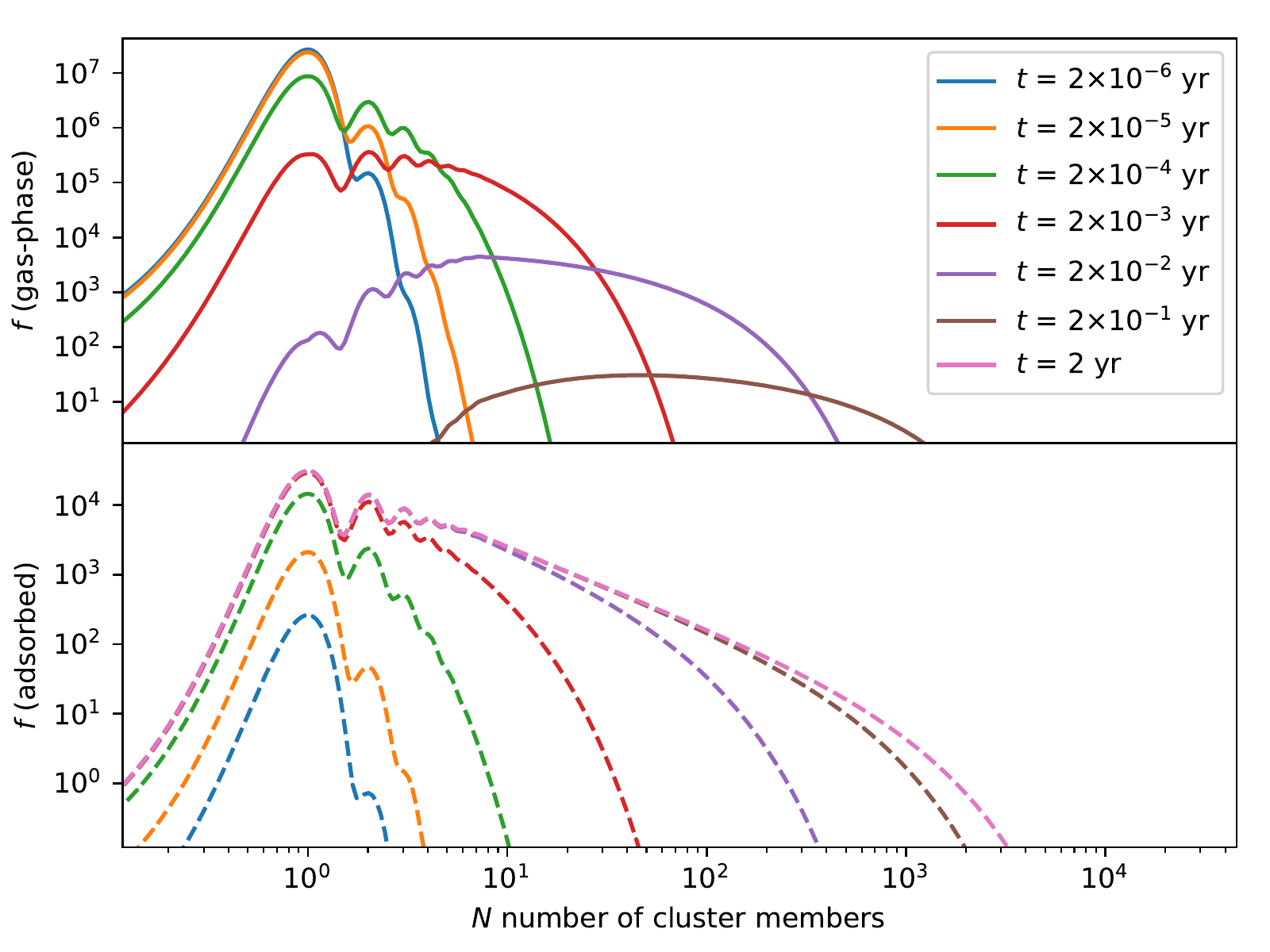}
    \caption{Similar to figure \ref{fig:PAHgrowthDust}, but the particle size distribution is shown. While most of the mass is adsorbed at the end of the coagulation, the most frequent cluster size is the monomer, which can only adsorb during the early stages of coagulation.}
    \label{fig:PAHgrowthDustNumbers}
\end{figure}

\begin{figure}
    \centering
    \includegraphics[width=1\linewidth]{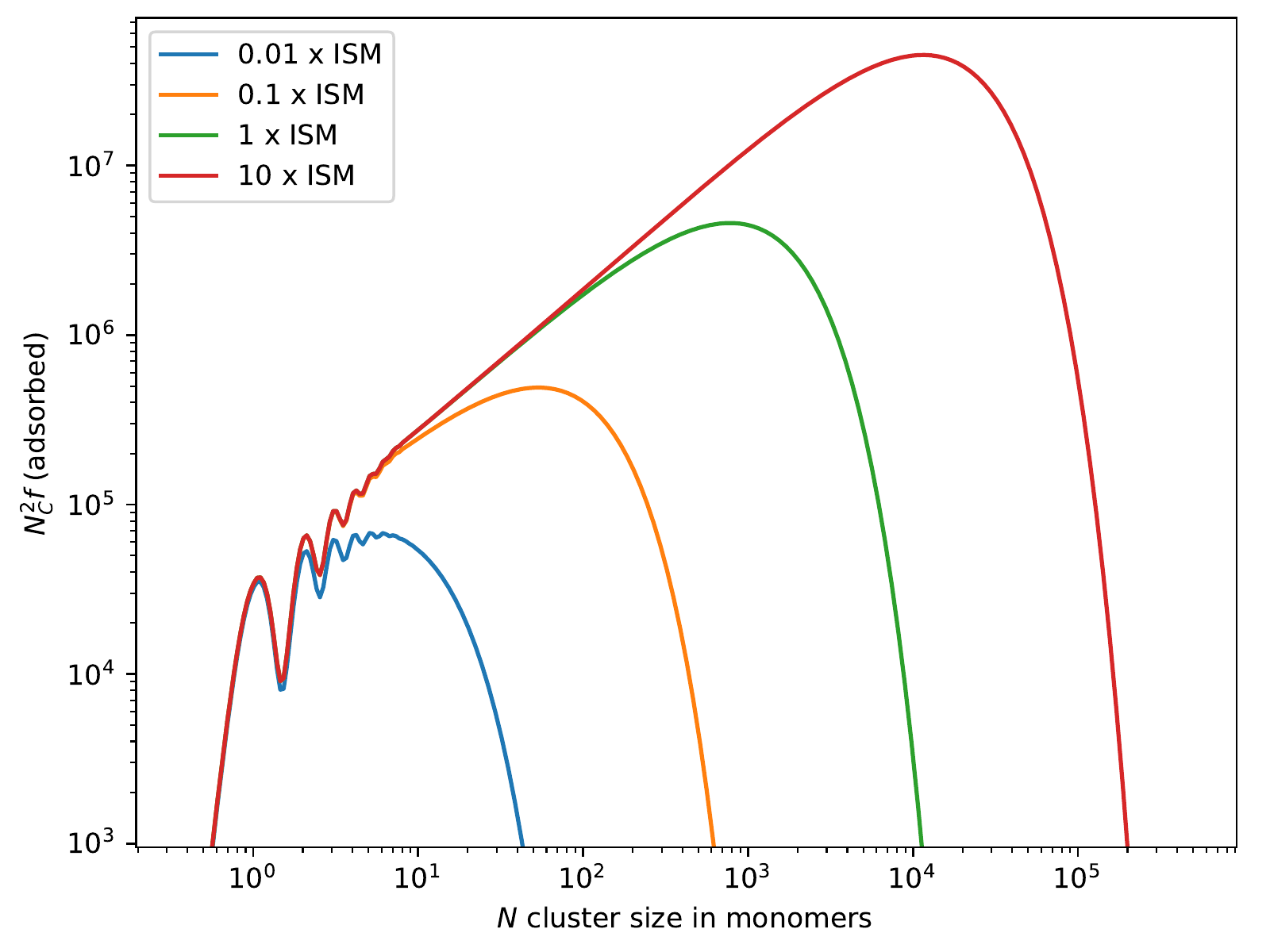}
    \caption{Size distribution of adsorbed PAH clusters as a function of cluster size for initial PAH abundances in the range 0.01--10 times the ISM abundance. For small clusters ($N\leq$10) a maximum abundance is reached with $0.1 \times$ ISM abundance that cannot be increased by further increasing the initial PAH abundance.}
    \label{fig:adsorption}
\end{figure}

In the following section we present the results of the PAH coagulation model with and without the presence of dust grains to show the effect on the PAH population.
We apply our Monte Carlo method to the standard disc model STD from Table \ref{tab:disc_models} and simulate the coagulation of coronene (C$_{24}$H$_{12}$) with an initial ISM abundance (ratio of hydrogen atoms to C atoms locked up in PAHs; C$_\text{PAH}$:H = $1.5\cdot10^{-5}$; \citealt{Tielens2008}).
We use $r=10$\,au and $\theta=0$ as representative of the entire coagulation layer since the ratio of PAH to dust collision cross-section is initially nearly constant (figure \ref{fig:sigma-sigma}).
Therefore, our coagulation simulations are universally applicable to the coagulation layer if the location is optically thick to the incident UV radiation to prevent dissociation of the clusters ($\tau_\text{UV}\gg 1$).
Additionally, the PAHs need to be uncharged as otherwise the electric repulsion force   inhibits the sticking of the PAHs.
We describe our PAH population by the number of cluster members ($N$ monomers = $N_\text{C}/N_\text{C,0}$) and give the corresponding PAH size distribution as $N_\text{C}^2 f$, so that the surface area under the curve is equivalent to the total number of C atoms in this size range (resp. proportional to the PAH mass).\\
\\Figure \ref{fig:PAHgrowth} shows the coagulation of PAH clusters without the presence of any dust grains.
Within the first few days after the start of the simulation, dimers, trimers, and quadrumers form whose respective peak width is given by the propagation of the initial conditions.
The further growth of the PAH clusters is rapid.
Within a year, clusters can form whose size corresponds to the size of very small carbonaceous grains (VSGs) ($a\approx 10\,\AA$ resp. $N_\text{C} \approx 10^4$ C atoms), as defined in the models by \citet{Desert1990} and \citet{Draine2001}, among others.
This growth continues until all PAHs are eventually locked up in very large PAH clusters (with $N \geq 10^{5}$ monomers).
Depending on the underlying PAH species, these very fast-forming clusters cannot be photodissociated at larger distances from Herbig stars (e.g. $r\geq40$\,au for circumcoronene) when they are transported into the photosphere as they exceed the critical cluster size \citep{Lange2021}.
Consequently, if the conditions are met for this scenario (lack of UV radiation and dust grains), we expect the formation of very large clusters that can hardly be broken up if transported to UV-rich environments, even for Herbig star discs.\\
\\However, a different picture emerges when adsorption on dust grains is considered with the same initial conditions as before (figure \ref{fig:PAHgrowthDust}).
The first phases of cluster growth are not disturbed by the dust grains as cluster growth is initially many times faster than adsorption.
However, this changes when the PAHs have grown to larger cluster sizes as the absolute number of clusters has decreased so much that the probability of PAH-PAH collisions becomes similar to PAH-dust particle collisions. 
Consequently, clusters themselves start to stick on dust grains, which further reduces the number of possible PAH-PAH collision partners.
From this point on, further cluster growth is mostly inhibited and thus the maximum cluster size is limited by the total available dust collision cross-section.
Based on the adsorbed distribution, most of the mass of PAHs is adsorbed as large clusters.
However, the most frequent adsorbed PAH size is still the monomer $f$ (figure \ref{fig:PAHgrowthDustNumbers}), and the occurrence decreases with size.
The adsorbed size distribution is identical for all dust particle sizes $a$ since the size of the PAHs is negligible compared to the dust particles. 
Moreover, the number of PAHs on a specific grain size is proportional to the total surface area $\sigma_\text{dust}(a)$ of the dust particle size range (see equation \eqref{eq:Q-}).
Generally, PAHs are more likely to freeze out on the smallest dust grains that are locally available.\\
\\Furthermore, we want to investigate how clustering is affected when the initial PAH abundance differs from the ISM abundance as we do not know the initial PAH content in the molecular cloud.
For comparison with the other model, we keep the dust grain population the same.
To do this, we vary the PAH abundance in the range 0.01-10 times the ISM abundance (C$_\text{PAH}$:H = $1.5\cdot10^{-5}$; \citealt{Tielens2008}) and compare the adsorbed PAH cluster distribution (figure \ref{fig:adsorption}).
If the PAH abundance is very low ($\leq0.01$ ISM abundance), the amount of adsorbed PAHs decreases with decreasing PAH abundance.
Due to the low number of PAH collision partners, the PAH clusters cannot grow very quickly and mainly small PAH clusters are formed that adsorb on the dust grains.
If the PAH abundance is sufficiently high ($\geq 0.1$ ISM abundance), the absolute number of adsorbed PAHs is limited for the smaller cluster sizes ($N$ $\leq$ 10) and does not   increase further, even though the initial PAH abundance has been increased.
The largest cluster size at which this maximum adsorbate abundance is reached depends on the initial PAH abundance and shifts with it to larger cluster sizes.\\
\\In general, the existence of this maximum adsorbate abundance can be explained by comparison of the adsorption rate and the clustering rate. 
A necessity for this is that PAH-PAH collisions are much more likely than PAH-dust particle collisions so that the PAH cluster growth is initially unaffected by the presence of the grains.
Then the rate at which PAH clusters collide with another PAH cluster $r_\text{PAH-PAH}$ is proportional to $f\left(N_\text{C}\right)^2$ (eq. \eqref{eq:coag}), while the rate of collisions with dust particles $r_\text{PAH-dust}$ is proportional to $f\left(N_\text{C}\right)$ (eq. \eqref{eq:Q-}).
Consequently, the  timescale $\tau$ on which all PAHs have grown is proportional to $f\left(N_\text{C}\right)^{-1}$ ($\tau = f\left(N_\text{C}\right) / r_\text{PAH-PAH}$\footnote{where $r_\text{PAH-PAH}$ is the PAH-PAH collision rate as before}$\propto f\left(N_\text{C}\right)/f\left(N_\text{C}\right)^2 = f\left(N_\text{C}\right)^{
-1}$). 
Accordingly, the absolute number of PAHs and clusters that can adsorb during this time is independent of the initial amount of PAH monomers $f\left(N_\text{C}\right)$ ($f_\text{adsorbed}\left(N_\text{C}\right) = \tau \cdot r_\text{PAH-dust} \propto f\left(N_\text{C}\right)^{-1} \cdot f\left(N_\text{C}\right) = \text{const}$).
Most importantly, the maximum adsorbed amount for small clusters ($N\leq10$) is already reached when the PAH abundance is equal to the ISM abundance which becomes relevant when we analyse the PAH desorption in the next section.\\
\\In summary, the presence of PAHs and dust grains in a strongly UV-shielded region of the protoplanetary disc inevitably leads to clustering of PAHs with subsequent adsorption of PAH monomers and PAH clusters onto dust grains.
Dust grains inhibit the further growth of PAH clusters, which otherwise would grow to carbonaceous nanograins. 
The adsorbed PAH clusters follow a size distribution whose number distribution is dominated by monomers and small clusters, but mass-wise is dominated by large clusters. 
Based on this, we  estimate in the next section how difficult it is to recover these clusters and monomers from the dust grains depending on their size.

\begin{figure*}[h]
    \begin{subfigure}
    \centering
    \includegraphics[width=0.49\linewidth]{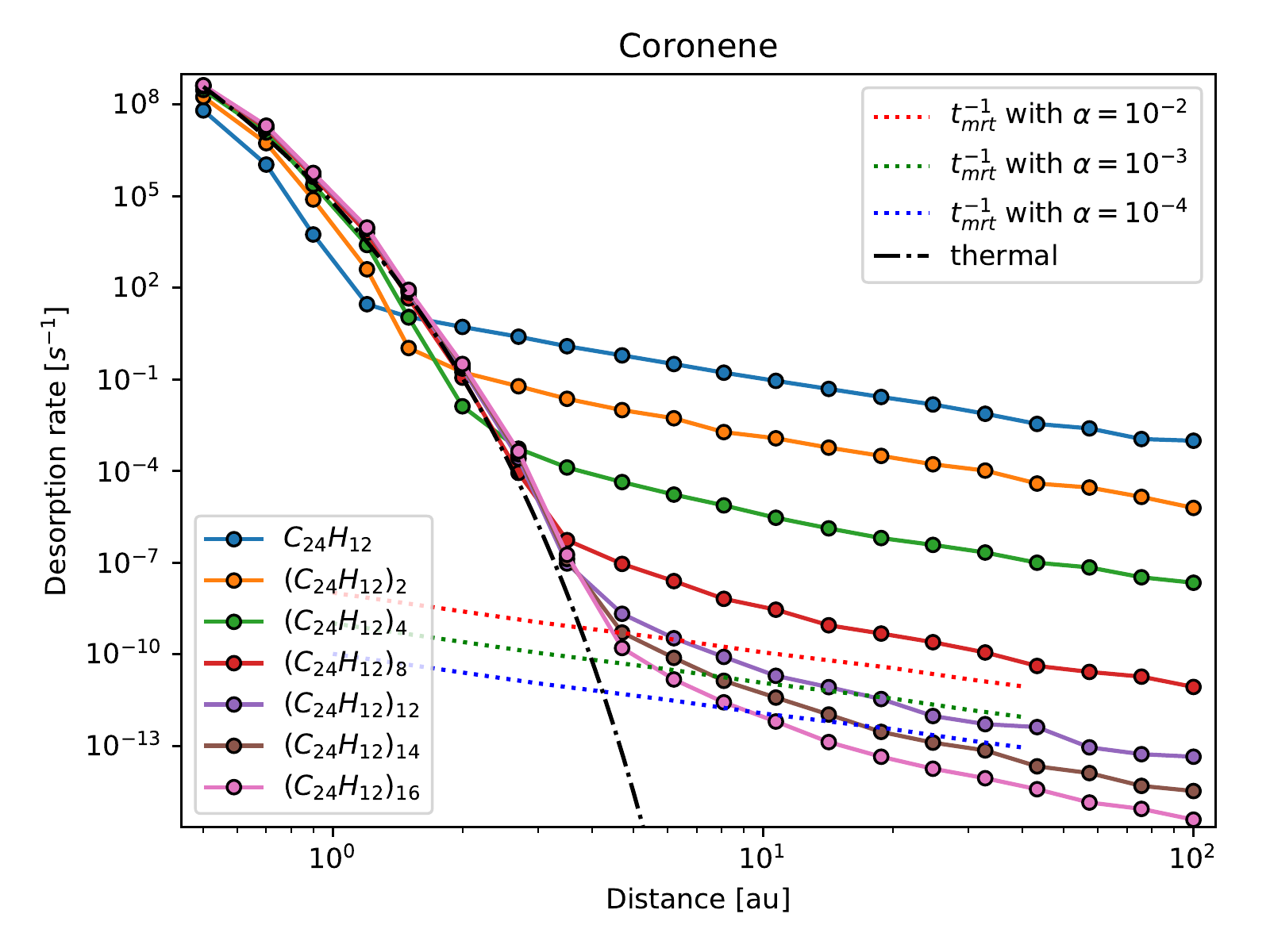}    
    \end{subfigure}
    \begin{subfigure}
    \centering
    \includegraphics[width=0.49\linewidth]{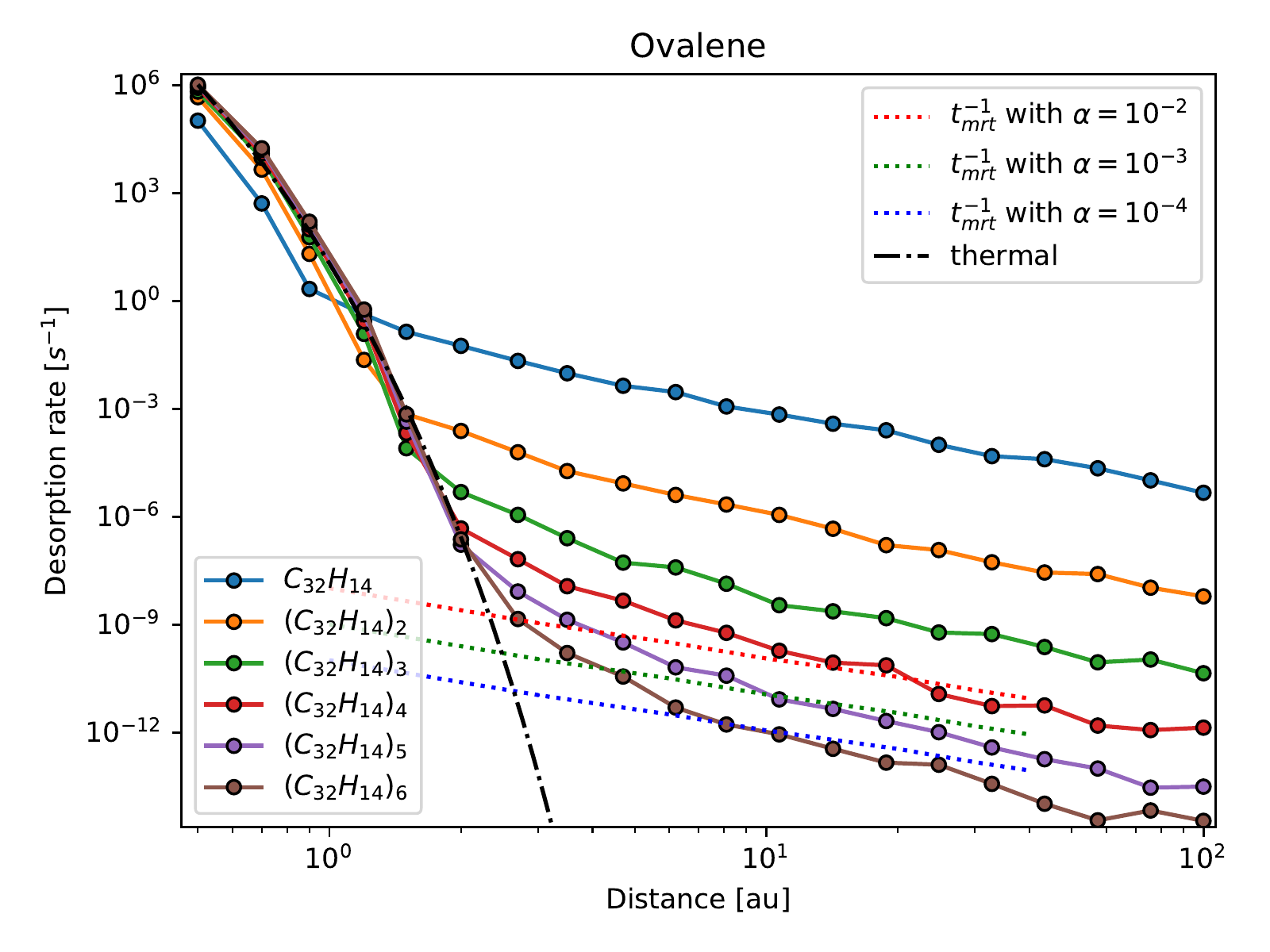}    
    \end{subfigure}\\
    \begin{subfigure}
    \centering
    \includegraphics[width=0.49\linewidth]{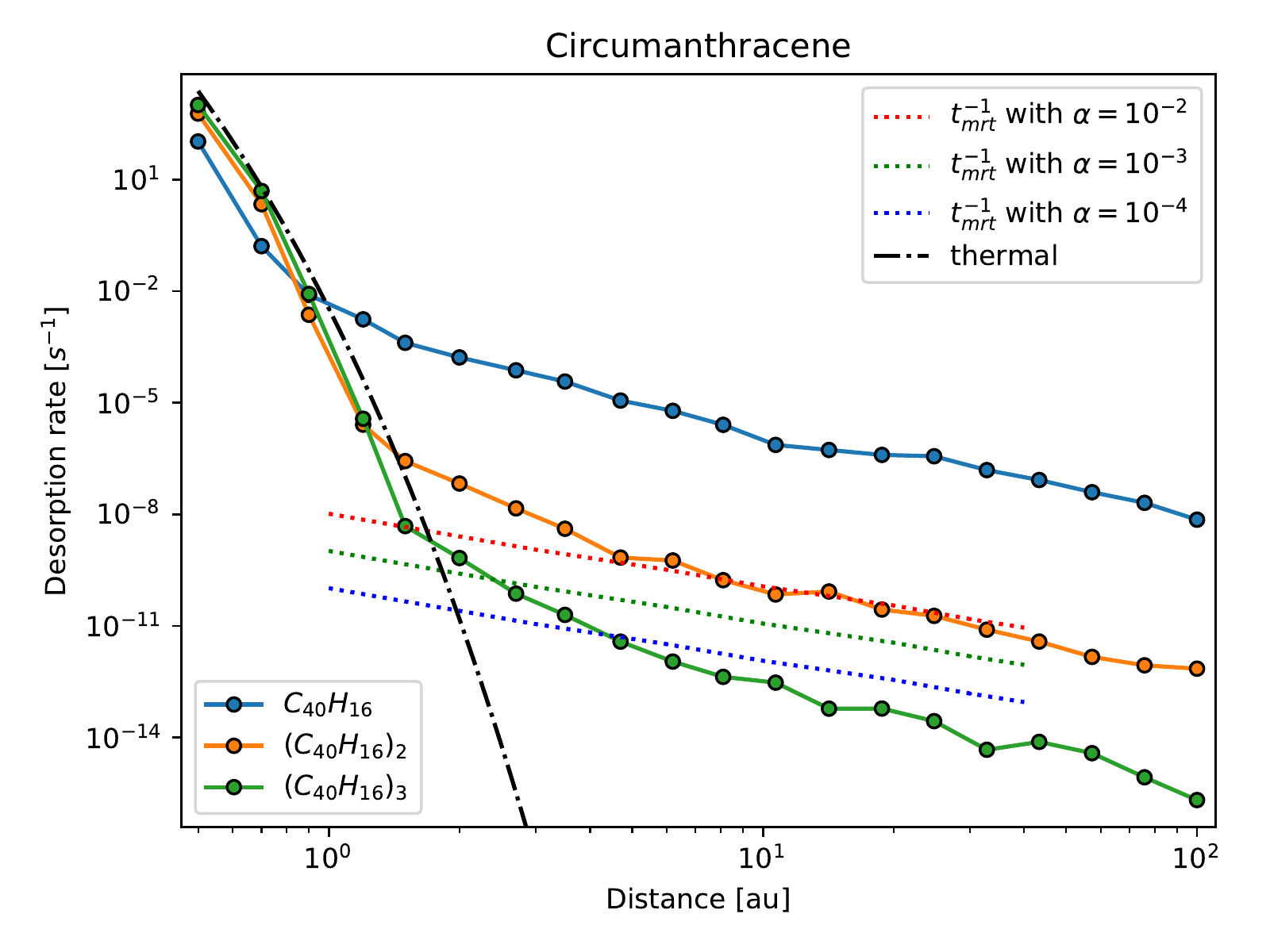}    
    \end{subfigure}
    \begin{subfigure}
        \centering
    \includegraphics[width=0.49\linewidth]{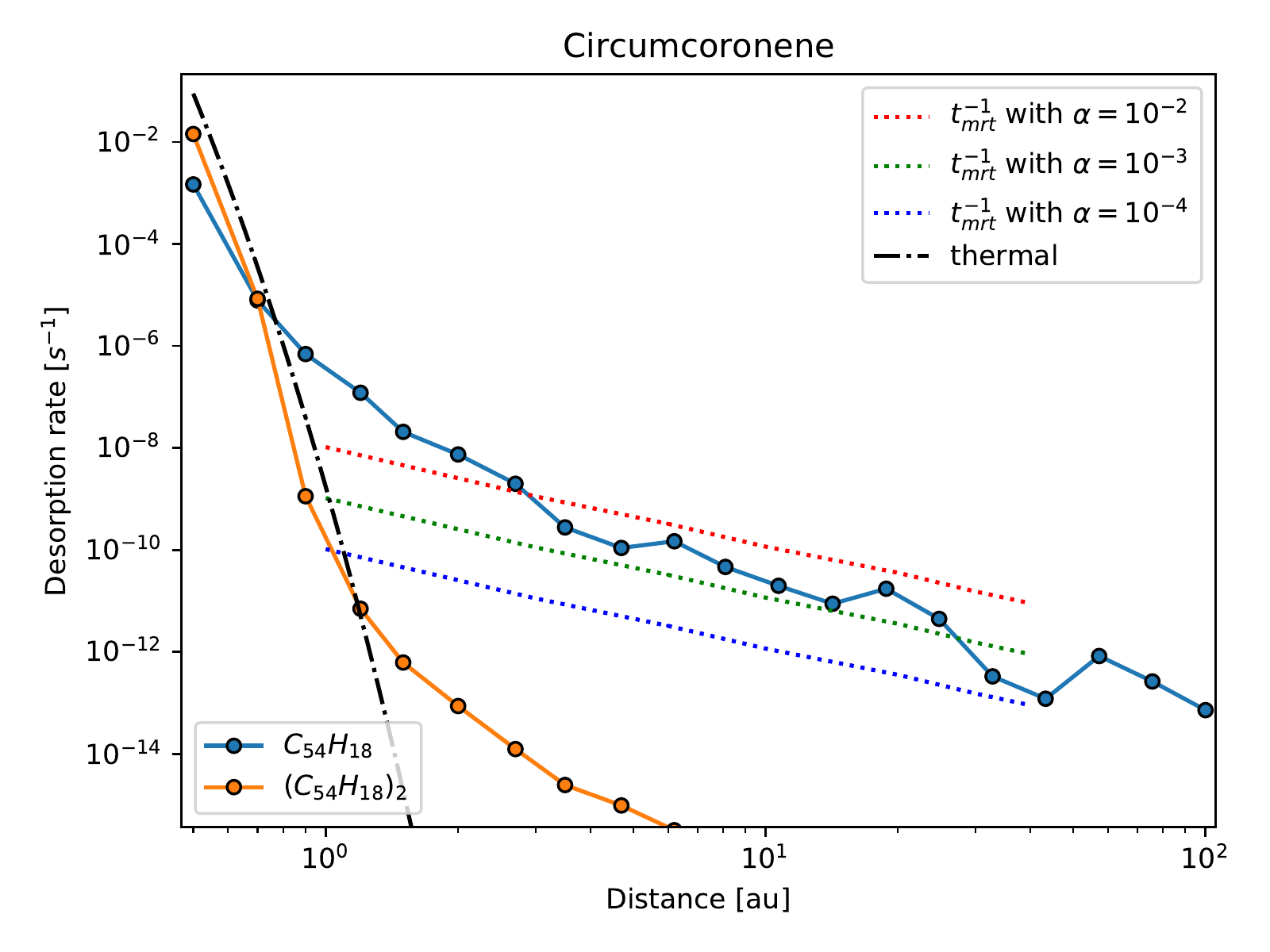}    
    \end{subfigure}

    \caption{Desorption rates in the Herbig disc STD photosphere calculated with the Monte Carlo model for increasing cluster sizes, radii, and PAH species. Shown are  thermal evaporation rates calculated from eq. \eqref{eq:arrhenius} and eq. \eqref{eq:T_dust} (dash-dotted lines) and  turbulent mean-residence time in the photosphere calculated in section \ref{sec:PAHdisc} (dotted lines). For increasing PAH size and cluster size, desorption becomes slower until   clusters are unlikely to desorb at all. PAH coagulation and adsorption on dust grains therefore lead to loss of gas-phase PAHs to the grains. More PAH species are available in figure \ref{fig:desorb_r_appendix}.}
    \label{fig:desorb_r}
\end{figure*}

%%%%%%%%%%%%%%%%%%%%%%%%%%%%%%%%%%%%%%%%%%%%%%%%%%%%%%%%%%%%%%%%%%%%%%%%%%%%%%

\subsection{Desorption of PAH clusters from grains}
\label{sect:PAH_desorb}
In this section we     investigate how fast the adsorbed clusters that formed in the coagulation layer can desorb again from the dust grains when they are brought into a UV-rich environment such as the photosphere.
For this purpose we use the Monte Carlo model described in section \ref{sec:desorp} to determine the energy probability distribution $G(E)$ of adsorbed PAH clusters to obtain the thermal and UV-assisted desorption rate.
We apply our model to clusters with a size of 1--16 monomers, and determine the desorption of the whole clusters from the grains as a function of the distance $r$ from the central star.
These calculations are performed for various PAH species whose monomers have between 24 (coronene) and 66 (circumovalene) C atoms and for which the absorption spectrum is available in the Cagliari theoretical spectral database of polycyclic aromatic hydrocarbons \citep{Malloci2007}.\\
\\Figure \ref{fig:desorb_r} shows the desorption rate of different cluster sizes for selected PAH species in the standard Herbig disc model STD from Table \ref{tab:disc_models} assuming that all PAHs are fully exposed to the stellar UV radiation. The calculations for additional PAH species can be found in figure \ref{fig:desorb_r_appendix} in Appendix \ref{app:desorp}.
The total desorption rate is shown as solid lines and the thermal evaporation rate of clusters at the equilibrium temperature of the grains is shown as a dash-dotted line (using Eqs. \eqref{eq:arrhenius}, \eqref{eq:Teff}, and \eqref{eq:T_dust}).
For all PAH species the desorption rate in the inner disc regions is completely dominated by the thermal evaporation of the clusters from the grain surface.
Recalling the rapid dissociation of PAH clusters in the photosphere \cite{Lange2021}, this high evaporation rate implies that PAHs are exclusively present as gas-phase monomers in the inner photosphere.
We note that the slightly lower thermal desorption rate for monomers and small clusters reflects the heat bath correction in eq. \eqref{eq:Teff}.\\
\\Increasing the distance to the central star, the thermal desorption rate decreases due to the lower equilibrium temperature of the dust grains and desorption through single UV photon absorption by the PAH cluster becomes the main cause of cluster evaporation.
The evaporation rate decreases approximately with the distance square $r^{-2}$ due to the reduction of the photon density, but only approximately since the dust temperature continues to decrease and the energy--temperature relation is not linear.
Overall, the larger a PAH cluster is, the less likely it is to evaporate from the grain surface as the peak excitation temperature decreases due to the larger heat capacity of the cluster.
The same effect occurs when a larger PAH species is considered, but the evaporation rate is further reduced by the fact that the grain-contact PAH monomer of the cluster is larger, resulting in a stronger van der Waals bond to the dust grain.
Hence, a coronene dimer (C$_{24}$H$_{12})_2$ detaches more quickly than a dicoronylene monomer (C$_{48}$H$_{20}$).\\
\\We \  compare the desorption rates of the clusters to the timescale these clusters would be exposed to the full stellar UV radiation in the protoplanetary disc turbulence model (see section \ref{sec:turbulence} and \ref{sec:PAHdisc}).
The inverse of the mean-residence times in the photosphere are plotted as dotted lines in figure \ref{fig:desorb_r}.
Provided the cluster is large enough that the desorption rate is lower than the inverse mean residence time in the photosphere, the cluster can remain adsorbed on the grains. 
Therefore, we define the size at which this condition is met as the critical cluster size.
We track the critical cluster size and use it in section \ref{sec:vertical_model} to determine how many PAHs can be recovered from the grains.
The derived critical cluster sizes at 10\,au can be found in Table \ref{tab:critsizes}.\\
\\In particular, we find that the evaporation rate of circumcoronene monomers is much lower than the exposure rate except for the inner disc regions.
Therefore, PAHs with more than 54 carbon atoms will stay frozen out on dust grains in most parts of the disc.
In view of these results, we can conclude that the coagulation and freeze-out of PAHs has two effects if these processed PAHs become exposed to stellar UV radiation.
The first effect is a permanent freeze-out in most parts of the disc as a large fraction of the PAH clusters will grow to sizes that cannot be recovered. Even though the effectiveness of this process depends on the PAH species, we expect that the gas-phase PAH abundance is smaller than the abundance in the ISM.
The second  is a size selection effect. Under the same conditions, smaller PAH species are more likely to be recovered than larger PAH species. Thus, we expect only small to medium gas-phase PAHs ($N_\text{C,0}\leq54$) in a vertically mixing disc, and the photosphere is expected to be dominated by the small PAH molecules. We note that we only consider a single PAH species in this model. Clusters of mixed PAH species are more difficult to treat because any PAH species could be the bonding molecule to the grain, which will affect the desorption rates.
Next, we want to quantify the strength of this depletion mechanism depending on the PAH abundance and dust grain population after one cycle of coagulation, freeze-out, and desorption.

%%%%%%%%%%%%%%%%%%%%%%%%%%%%%%%%%%%%%%%%%%%%%%%%%%%%%%%%%%%%%%%%%%%%%%%%%%%%%%
%%%%%%%%%%%%%%%%%%%%%%%%%%%%%%%%%%%%%%%%%%%%%%%%%%%%%%%%%%%%%%%%%%%%%%%%%%%%%%
%%%%%%%%%%%%%%%%%%%%%%%%%%%%%%%%%%%%%%%%%%%%%%%%%%%%%%%%%%%%%%%%%%%%%%%%%%%%%%
\subsection{The structure of the PAH disc}
\label{sec:PAHdisc}
\begin{figure}
    \centering
    \includegraphics[width=\linewidth]{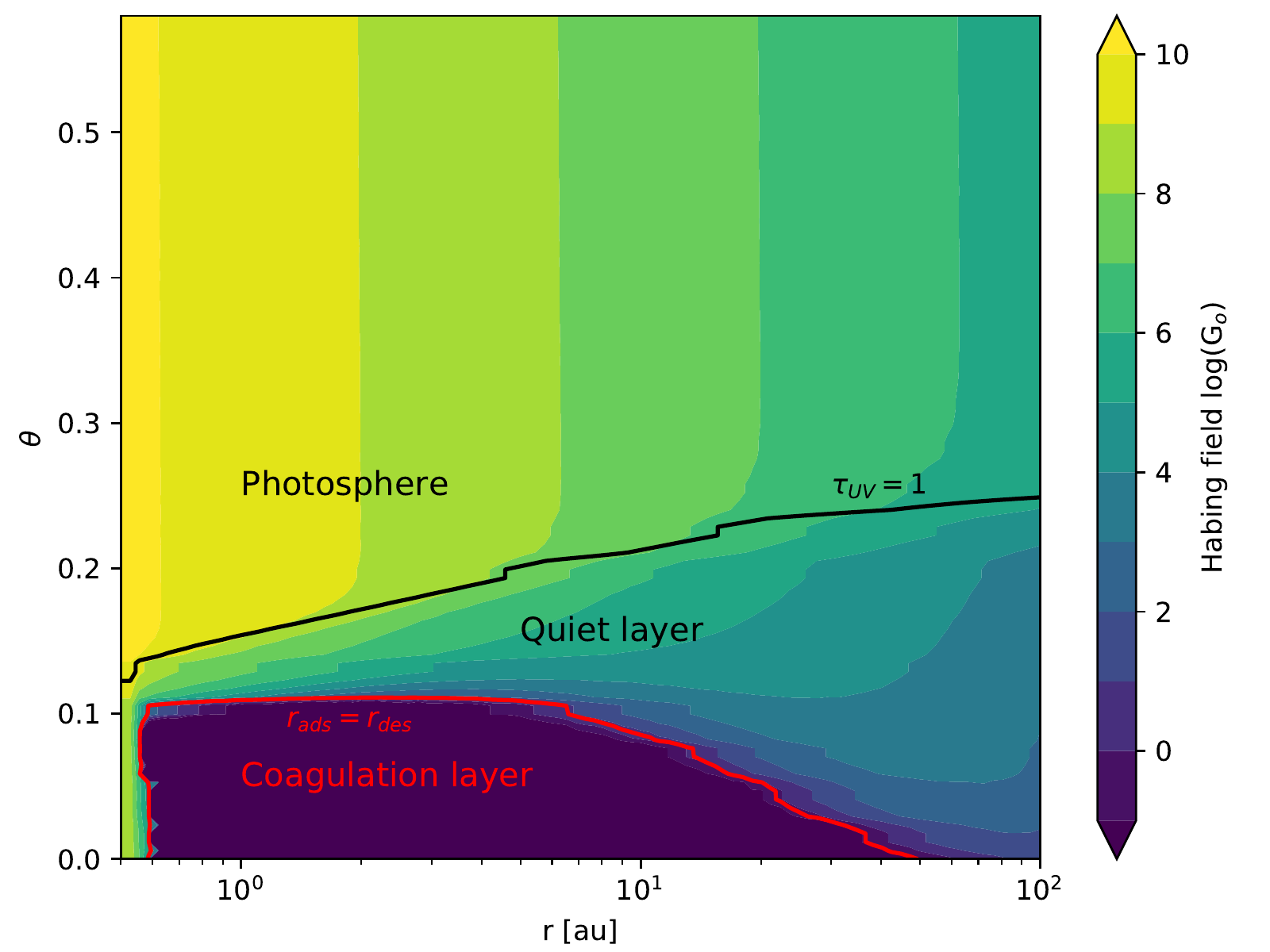}
    \caption{Intensity of the mean UV radiation field $G_0$. The black line indicates the $\tau_\text{UV}=1$ line, and the red line shows where the monomer desorption rate is equal to the monomer adsorption. In the photosphere the small PAH clusters quickly desorb and dissociate into monomers. In the coagulation all gas-phase PAHs coagulate and adsorb on dust grains, while in the quiet layer monomers, clusters, and PAH-carrying grains can co-exist.}
    \label{fig:layererd_disk}
\end{figure}

We use the radiative transfer code RADMC-3D\footnote{\url{https://www.ita.uni-heidelberg.de/~dullemond/software/radmc-3d/contributions.php}} and supplementary tool radmc3dPy\footnote{ \url{https://www.ita.uni-heidelberg.de/~dullemond/software/radmc-3d/manual_rmcpy/index.html}} with our disc model STD  to estimate the expected UV field between the midplane and the photosphere as this region is attenuated.
For the opacities of the dust grains, we use the tool OpTool \citep{Dominik2021} with the standard DIANA opacities \citep{Woitke2016}.
We measure the UV field in the range 6--13.5\,eV in units of the mean interstellar UV field $G_0 = 5.33 \cdot 10^{-14}$\,erg cm$^{-3}$ \citep{Habing1968}.
As desorption occurs only through single photon events, we can use a power law description similar to that in  \cite{Lange2021} and scale the desorption rate by $G_0$ assuming that the attenuation from the dust grains does not change significantly between 6 and 13.5\,eV.
Figure \ref{fig:layererd_disk} shows the strength of the UV field $G_0$ in the given disc model. The $\tau_{UV}=1$ line (\textit{black}) where $G_0$ has dropped by $1/e$ and the coagulation front (\textit{red}) where the adsorption and desorption rate of PAH monomers are equal.
We find that the disk is vertically structured in three physically layers: the photosphere, the quiet layer, and the coagulation layer.\\
\\In the photosphere, the PAH molecules are fully exposed to stellar radiation.
Small clusters can evaporate from grains on the  timescales given in section \ref{sec:coag}, and all clusters that desorb can also be dissociated on a faster timescale \cite{Lange2021}.
Therefore in this layer, PAHs exist mainly as gas-phase monomers and large adsorbed clusters on grains.
As gas-phase PAHs move into the quiet layer, the UV field weakens and the density increases.
However, as shown by \citet{Thi2019}, PAHs are the major carrier of negative charges at these heights.
Therefore, we do not expect cluster formation through collisions to be effective because of the electric repulsion force between two molecules. 
At the same time, small adsorbed clusters coming from the midplane stay adsorbed on the grains as the UV field is significantly weaker than in the photosphere.
Hence, gas-phase monomers and small adsorbed cluster rarely desorb, cluster, or adsorb.
For simplicity, we neglect all these processes here as a full model tracking the state of all PAHs is required.
The upper boundary of the coagulation layer, the coagulation front, is located where the UV field equals the interstellar field ($G_0 \approx 1$). 
At this location, the conditions match as PAH charging is not significant any longer, due to the lack of UV photons, and the densities are high enough that coagulation and adsorption are faster than their UV-driven counter processes. 
\begin{table}[]
    \caption{Critical cluster sizes at 10\,au for a turbulence value of $\alpha = 10^{-3}$ with corresponding binding energies $E_\text{A}$.}
    \label{tab:critsizes}
    \centering
    \begin{tabular}{l|c|c}
    PAH species & critical size at 10\,au & $E_\text{A}$\\
    \hline
         Coronene & $\left(\text{C}_{24}\text{H}_{12}\right)_{13}$ & 1.4\,eV \\
         Bisanthene & $\left(\text{C}_{28}\text{H}_{14}\right)_{7}$ & 1.6\,eV \\
         Ovalene & $\left(\text{C}_{32}\text{H}_{14}\right)_{5}$ & 1.8\,eV \\
         Tetrabenzocoronene & $\left(\text{C}_{36}\text{H}_{16}\right)_{3}$ & 2.0\,eV \\
         Circumanthracene & $\left(\text{C}_{40}\text{H}_{16}\right)_{2}$ & 2.2\,eV \\
         Dicoronylene & $\left(\text{C}_{48}\text{H}_{20}\right)_{1}$ & 2.6\,eV \\
         Circumcoronene & $\left(\text{C}_{54}\text{H}_{18}\right)_{1}$ & 2.9\,eV \\
         Circumovalene & $\left(\text{C}_{66}\text{H}_{20}\right)_{0}$ & 3.5\,eV
    \end{tabular}
    
    \label{tab:PAH_props}
\end{table}

\subsection{Recoverable PAH abundance}
\begin{figure}
    \centering
    \includegraphics[width=0.99\linewidth]{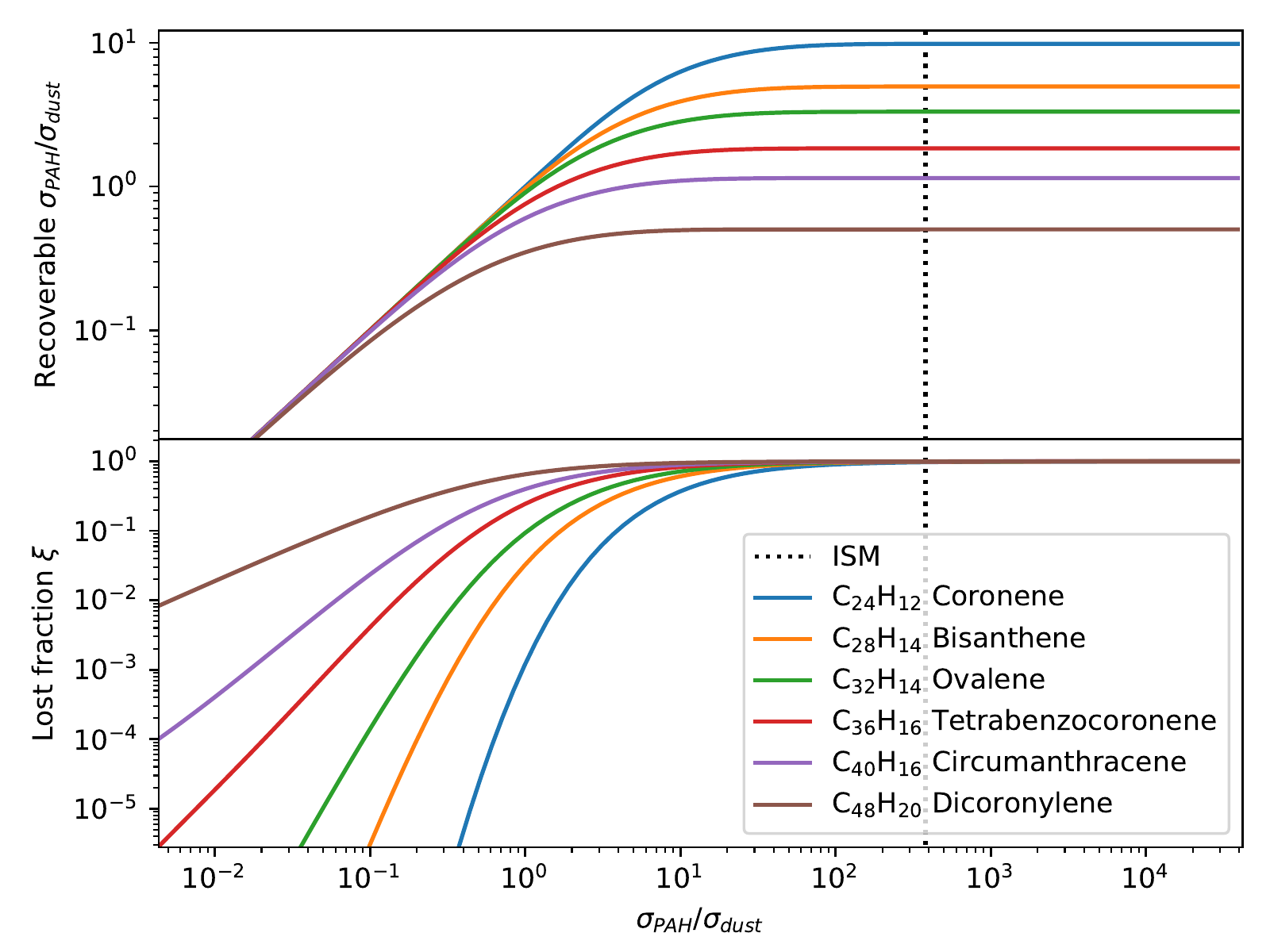}
    \caption{Recoverable $\sigma_\text{PAH}/\sigma_\text{dust}$ ratio and lost fraction of PAHs after one cycle of clustering for different PAH species and different initial $\sigma_\text{PAH}$ using the standard dust model. The dotted black line assumes a standard ISM abundance of PAHs. For PAH abundances higher than $\sigma_\text{PAH}/\sigma_\text{dust}\geq 30$ ($\approx 0.1  \times$ ISM) the maximum abundance is reached (section \ref{sec:res_coag}) and information on the initial PAH abundance is lost. For $\sigma_\text{PAH}/\sigma_\text{dust}< 1$, almost all PAHs can be recovered because adsorption is much faster than coagulation and processing has no effect on the PAH abundance any longer as all PAHs are preserved.}
    \label{fig:PAH_lost_recov}
\end{figure}

From the results of the previous sections, we     derive the number of PAHs that can re-enter the gas phase (hereafter recoverable PAHs) as these are the major emitters of the infrared bands through their strong temperature fluctuations.
In order to be recoverable, a PAH cluster must be able to desorb from the grain and dissociate within the mean residence time in the photosphere.
For simplicity, we approximate the total time to become a monomer from a cluster frozen on a grain by the desorption timescale as this is the longest timescale,
otherwise a full model tracing the PAHs through the photosphere and quiet layer would be necessary to estimate the time for dissociation and desorption.
To obtain a generally applicable result we parametrise the PAH and dust population by the total available collision cross-section per unit area in the PAHs $\sigma_\text{PAH}$ compared to the total collision cross-section per unit area of the dust $\sigma_\text{dust}$.
This is possible as the competition between clustering and freeze-out determines the fraction of PAHs below the critical cluster size that can desorb again.\\
\\To obtain the recoverable fraction of adsorbed PAH clusters, we run our coagulation model with varying initial PAH abundances and determine the adsorbed cluster size distribution $f(N_\text{C})$. 
According to the calculated critical cluster sizes, we then calculate what fraction of PAHs is recoverable from the dust grains by integration of all clusters that are smaller than the critical cluster size
\begin{equation}
    \xi = \frac{f_\text{tot, recov}}{f_\text{tot}} = \frac{\int_0^{N_\text{C,crit}} f dN_\text{C}}{\int_0^\infty f dN_\text{C}} \text{.}
    \label{eq:fracloss}
\end{equation}
In our case, this ratio can be used generally throughout the disc as $\sigma_\text{PAH}/\sigma_\text{dust}$ is constant (see figure \ref{fig:FNcomparison}).
However, when a different dust population is considered and even allowed to evolve, then $\sigma_\text{dust}$ might not be suitable any more as the surface area can vary with height and in time.\\
\\Figure \ref{fig:PAH_lost_recov} shows the recoverable ratio $\sigma_\text{PAH}/\sigma_\text{dust}$ and the lost fraction $\xi$ at 10\,au after one cycle of coagulation and freeze-out as a function of the initial surface ratio of PAHs to dust grains for different PAH species.
The typical ISM PAH abundance is shown as a dotted line.
If $\sigma_\text{PAH}/\sigma_\text{dust} \leq 10^2$ (with the  STD model this equals 0.1 times ISM abundance of PAHs), the recoverable $\sigma_\text{PAH}/\sigma_\text{dust}$ ratio reaches a plateau, which is the maximum recoverable fraction from the dust grains. 
As explained in chapter \ref{sec:res_coag}, this is due to the much faster growth of clusters compared to adsorption on dust particles, so that the absolute amount of PAHs that can be adsorbed is constant and independent of the PAH abundance. 
As a consequence after one cycle, any information on the initial $\sigma_\text{PAH}/\sigma_\text{dust}$ is lost, as long as $\sigma_\text{PAH}/\sigma_\text{dust}$ is large enough.
Hence, no higher ratio than the maximum recoverable $\sigma_\text{PAH}/\sigma_\text{dust}$ can be expected when PAHs have processed through one cycle.\\
\\In contrast, if the PAH abundance is very low at the beginning, almost all PAHs can be recovered due to slow cluster growth and only a small fraction of PAHs is lost in a coagulation-adsorption cycle. The lower $\sigma_\text{PAH}/\sigma_\text{dust}$ becomes, the smaller the fraction that is lost per cycle as only small clusters form during the coagulation.
Thus, when PAHs are processed in many cycles, the loss of PAHs over time slows down until the lost fraction is negligible and the PAH abundance is effectively constant.
For this reason, we want to estimate how many PAHs are lost through adsorption within the typical protoplanetary disc life time.


%%%%%%%%%%%%%%%%%%%%%%%%%%%%%%%%%%%%%%%%%%%%%%%%%%%%%%%%%%%%%%%%%%%%%%%%%%%%%%
%%%%%%%%%%%%%%%%%%%%%%%%%%%%%%%%%%%%%%%%%%%%%%%%%%%%%%%%%%%%%%%%%%%%%%%%%%%%%%
%%%%%%%%%%%%%%%%%%%%%%%%%%%%%%%%%%%%%%%%%%%%%%%%%%%%%%%%%%%%%%%%%%%%%%%%%%%%%%

\subsection{A simple turbulent PAH model}
\label{sec:vertical_model}
\begin{figure}
    \centering
    \includegraphics[width=1\linewidth]{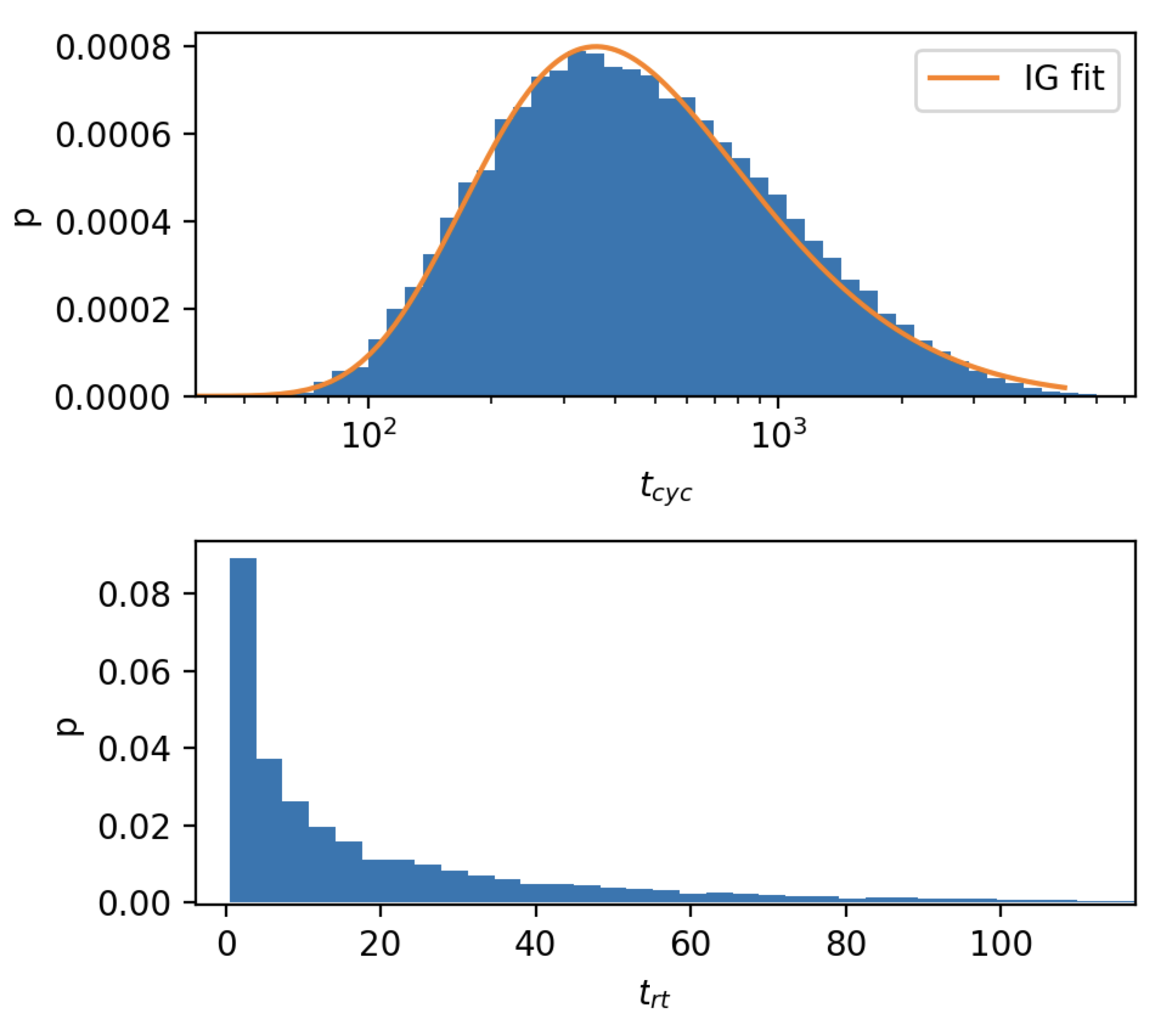}
    \caption{Distribution of measured cycle times $t_\text{cyc}$ and residence time in the photosphere $t_\text{rt}$ with a Monte Carlo framework at 40\,au with $\alpha = 10^{-2}$. The cycle time can be approximated with an inverse-Gaussian distribution. The mean-residence time is dominated by short timescales corresponding to an immediate mixing down into the quiet layer after reaching the photosphere.}
    \label{fig:t_dist}
\end{figure}

To determine the time evolution of the PAH abundance in a protoplanetary disc through persistent processing by clustering and adsorption, we   consider a simple 1D vertical model.
We look at a vertical column and assume that gas-phase PAHs and the smallest dust grains can be transported up and down the disc by vertical eddies following the equations presented in section \ref{sec:turbulence}.
We then follow a coupled particle with a Monte Carlo approach through the disc to determine the mean residence time $t_\text{mrt}$ in the disc photosphere and the mean cycle time $t_\text{mcyc}$.
Starting with the entry into the coagulation layer, the mean cycle time is the time needed to re-enter the coagulation layer provided that the photosphere has been reached before.
For the radial distance dependent boundaries of the photosphere and coagulation layer we use the results from the radiative transfer model (figure \ref{fig:layererd_disk}).\\
\\Figure \ref{fig:t_dist} shows the distribution of $t_\text{cyc}$ (\textit{upper panel}) and $t_\text{rt}$ (\textit{lower panel}) for the $\alpha=10^{-2}$ case at 40\,au as an example.
The mean cycle time can be estimated through an inverse-Gauss function, as the travel between photosphere and coagulation layer can be described as the first passage problem solved by \citet{Smoluchowsky1915}.
However, a slight deviation is expected as the classical random walk uses a 50\% probability for each step, while in the turbulence model the probabilities are slightly dependent on height.
We find that both distributions are heavily skewed to the left and dominated by smaller values.
Given that the number of cycles that can be completed in a typical protoplanetary disc life time of 2.5\,Myr \citep{Mamajek2009} is low, the averages of $t_\text{cyc}$ and $t_\text{mrt}$ should be used with care;  a detailed model tracking PAHs and its processing at each height would be more appropriate.
However, for the sake of simplicity, we use the mean values in this model.\\
\begin{figure*}
    \centering
    \includegraphics[width=0.99\linewidth]{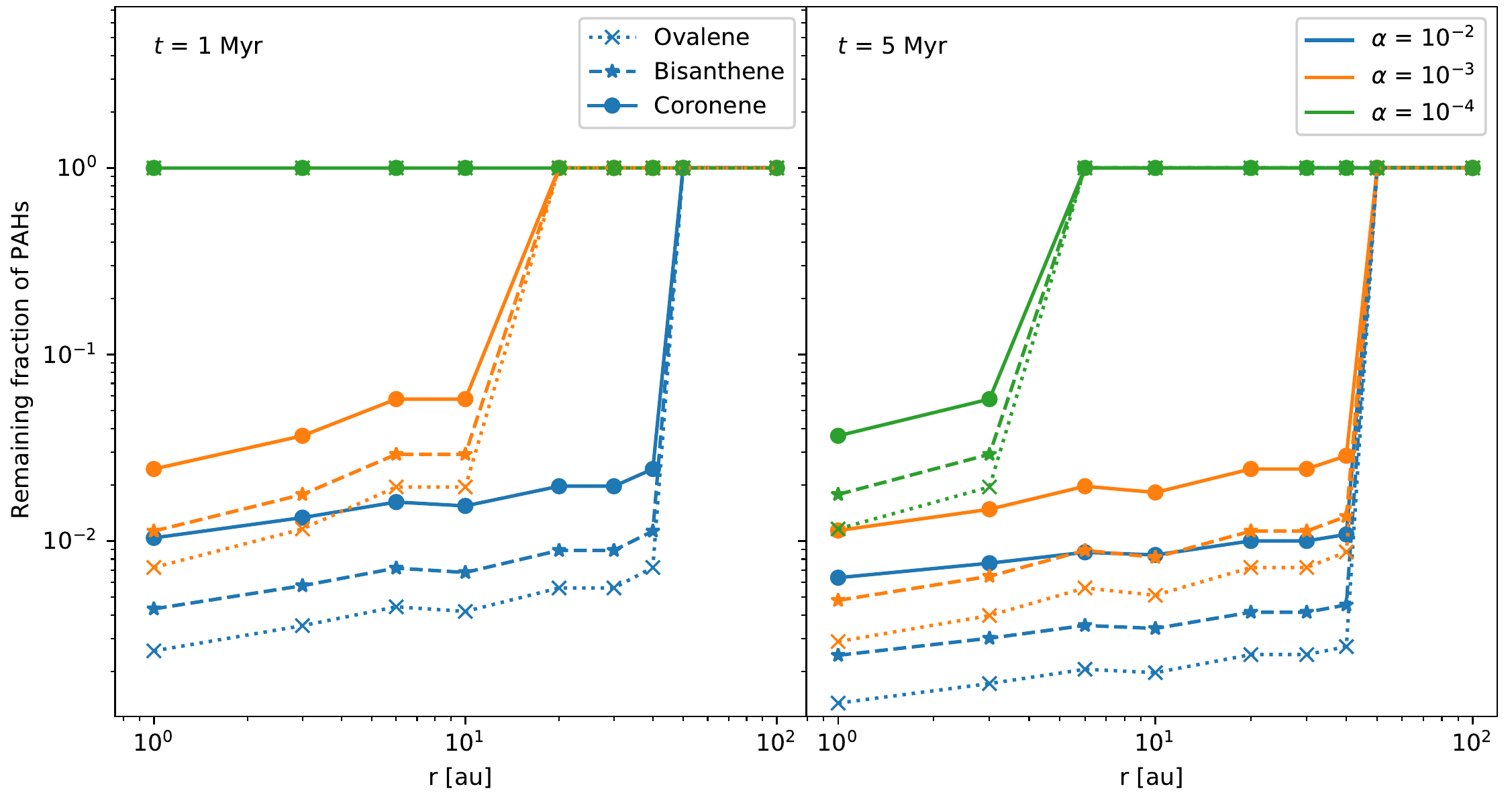}
    \caption{Turbulent depletion of PAHs through processing by clustering, adsorption, desorption, and photodissociation in the disc at 1\,Myr and 5\,Myr after formation. The calculations were performed for coronene C$_{24}$H$_{12}$, bisanthene C$_{28}$H$_{14}$, and ovalene C$_{32}$H$_{14}$. Strong ($\alpha = 10^{-2}$) and intermediate turbulence ($\alpha = 10^{-3}$) cause a depletion of recoverable PAHs by a factor of 50--500 compared to the initial ISM abundance in the model STD. At distances larger than 50\,au we do not find a layer where adsorption and therefore a full cycle can happen. For larger species than ovalene, the depletion increases when   the PAH species is larger.}
    \label{fig:cycle_evolution}
\end{figure*}
\\We calculate these timescales at all radial positions between $r_\text{min}$ and $r_\text{max}$.
Then we can estimate what the largest recoverable cluster size is (see section \ref{sec:desorp}) and determine the loss of unrecoverable PAHs deposited as clusters on grains.
We assume that after one mean cycle time $t_\text{cyc}$ all PAHs have undergone one complete cycle of mixing from the coagulation layer to the photosphere and back.
Then we can estimate the average PAH loss in one cycle with equation \eqref{eq:fracloss} by obtaining the recoverable fraction of PAHs through the relative collision cross-section ratio $\sigma_\text{PAH}/\sigma_\text{dust}$.
Figure \ref{fig:cycle_evolution} shows the depletion of gas phase PAHs through turbulent processes after 1\,Myr and 5\,Myr as a function of distance from the central star, turbulent strength, and PAH molecule.
For a strong mixing disc ($\alpha \approx 10^{-2}$), we expect a depletion of at least a factor 100 in the inner disc regions independent of the sampled PAH species as many cycles can be completed.
With a lower $\alpha$ fewer cycles can be achieved in the same time span.
For a very weakly mixing disc with $\alpha = 10^{-4}$, no cycle can be achieved in 1\,Myr.
Beyond 50\,au our disc model is not optically thick enough (we require $G_0 \approx 1$) to allow for the formation of the coagulation layer.
Hence, we do not expect turbulent processing in the outer parts of the disc as no new clusters can form.
After 5\,Myr, a depletion also occurs in the very inner regions ($\approx5$\,au), while in the $\alpha = 10^{-2}$ and $\alpha = 10^{-3}$ cases the depletion is stronger and more extended.\\
\\Despite the clear $\alpha$ turbulence dependence, we note that the used dust model from \citet{Fromang2007} was constructed to fit a $\alpha \approx 10^{-3}$ disc below four gas pressure scale heights, which we have used for all three $\alpha$ cases.
Therefore, with a different $\alpha$ the vertical distribution of the grains will change \citep[e.g.][]{Dullemond2004} as well as the height of the coagulation front and photosphere, which are tightly coupled to the optical thickness caused by the dust grains.
These effects were not considered in this study.

%%%%%%%%%%%%%%%%%%%%%%%%%%%%%%%%%%%%%%%%%%%%%%%%%%%%%%%%%%%%%%%%%%%%%%%%%%%%%%
%%%%%%%%%%%%%%%%%%%%%%%%%%%%%%%%%%%%%%%%%%%%%%%%%%%%%%%%%%%%%%%%%%%%%%%%%%%%%%
%%%%%%%%%%%%%%%%%%%%%%%%%%%%%%%%%%%%%%%%%%%%%%%%%%%%%%%%%%%%%%%%%%%%%%%%%%%%%%
%%%%%%%%%%%%%%%%%%%%%%%%%%%%%%%%%%%%%%%%%%%%%%%%%%%%%%%%%%%%%%%%%%%%%%%%%%%%%%
%%%%%%%%%%%%%%%%%%%%%%%%%%%%%%%%%%%%%%%%%%%%%%%%%%%%%%%%%%%%%%%%%%%%%%%%%%%%%%

\section{Discussion}
\label{sec:discussion}
\subsection{Initial conditions and disc infall}
In our study we  assume that the PAH abundance at the beginning of the disc stage is comparable to the value found in the ISM.
However, we  neglect the evolution of PAHs during the infall stage of the disc starting from the molecular cloud phase.
As the local density increases during the infall, coagulation of PAHs and adsorption on pristine dust grains might occur.
In order for the PAHs to not be dissociated or desorbed immediately, the local UV field must match the $G_0 \approx 1$ condition.
Whether this is possible is questionable as the local UV field is driven by the external UV field and the evolving and embedded protostar.
In the case that these conditions can be achieved in the infall stage, then PAHs will already enter the disc stage as adsorbed clusters.
As the first coagulation cycle already reduces the recoverable fraction to  1/30  of the initial value (e.g. for coronene), the disc would lose a substantial fraction of gas-phase PAHs.
Consequently, if all PAHs were   coagulated and adsorbed once already before the disc stage, then the outer regions of the disc ($\geq 50$\,au) would show a depletion of PAHs and the additional depletion through turbulent processing in the inner disc would have a much smaller effect.
Hence, the depletion of PAHs in the inner disc by continuous vertical mixing would be much less prominent or not even existent.\\
\\However, in our Herbig disc model we only take into account photons with less than 13.6\,eV and do not consider the hard photons   expected for T\,Tauri star discs originating from stellar accretion.
The interaction of individual PAH molecules in a turbulent discs with such hard photons has been investigated by \citet{Siebenmorgen2010} and \citet{Siebenmorgen2012}.
The authors conclude that typical X-ray luminosities can be an efficient mechanism to destroy individual PAH molecules in the disc.
However, as PAH clusters have a higher heat capacity, and frozen-out PAHs can transfer heat much faster than the IR-emission timescale considered in these works, the interaction of hard photons with frozen-out PAHs and PAH clusters should be considered in a future study.

%%%%%%%%%%%%%%%%%%%%%%%%%%%%%%%%%%%%%%%%%%%%%%%%%%%%%%%%%%%%%%%%%%%%%%%%%%%%%%
%%%%%%%%%%%%%%%%%%%%%%%%%%%%%%%%%%%%%%%%%%%%%%%%%%%%%%%%%%%%%%%%%%%%%%%%%%%%%%

\subsection{Typical PAH size and abundance in protoplanetary discs}
Our results from the desorption model strongly constrain the expected PAH size in protoplanetary discs.
Even a single cycle of clustering and adsorption leads to a size selection effect, where the expected abundance of smaller PAHs (coronene) is a factor of 3 higher than the abundance of medium-sized PAHs (ovalene) if initially they had the same abundance.
Additionally, PAHs that are larger than circumcoronene (C$_{54}$H$_{18}$) cannot be desorbed, even as a monomer.
Therefore, according to our model, we expect a distribution of small PAHs with less than 66 C atoms to be the dominant gas-phase PAH species and, due to their low heat capacity, also the dominant emitters of the IR features as the temperature of the adsorbed PAHs is similar to the grain temperature.
Our result agrees with the study of \citet{Seok2017}, where the authors fit emission models to protoplanetary disc spectra. 
Most of their best fit models predict PAH sizes between 3.5 and 6\,$\mu$m, which translate to PAH monomer sizes between 15 and 44 C atoms, where the upper size limit is a bit smaller than our largest desorbable PAH circumcoronene (C$_{54}$H$_{18}$).
Further, \citet{Seok2017} report a small negative correlation between PAH size and stellar age, which the authors expect to be outgassing of comets or grinding of planetesimals.
However, as the depletion per cycle for larger PAH species in our model is greater than for smaller PAH species and locally all PAHs experience the same mixing, we expect a correlation between PAH size and disc age by turbulent processing as well.\\
\\Typically, the common assumption is that the monomer size of a protoplanetary disk PAH is $N_\text{C,0} = 100$ C atoms \citep{Siebenmorgen2010,Maaskant2014}.
Similarly to the observationally constrained size by \citet{Seok2017}, we   favour PAH species with less than 66 C atoms as the typical disc PAH, as larger species can only desorb thermally in the disc. Therefore, we propose a significant size difference for PAHs between the interstellar medium and protoplanetary discs.
The ISM estimates for astronomical PAHs are usually based on observations of the ISM \citep{Allamandola1989} and nebulae like NGC7023 \citep{Croiset2016}, which emphasises the difference between the interstellar PAH population and disc PAHs.
Unfortunately, the ISM estimates are derived from temperature fluctuations of the PAHs using the direct link between heat capacity and size.
This size estimate is hardly possible in protoplanetary discs without spatially resolved images at multiple PAH emission wavelengths and a disc model to estimate emission heights and optical depth.
The James Webb Space Telescope (JWST) with its instruments NIRcam and MIRI will be one step to understand the difference between ISM PAHs and disc PAHs.
With its filters at 3.3\,$\mu$m and 11.3\,$\mu$m, we can start to analyse possible PAH structures in protoplanetary discs.\\
\\We also expect a depletion of PAHs in protoplanetary discs because of the processing of PAHs through coagulation, adsorption, desorption, and dissociation during their life times.
Even one cycle can lower the observable PAH abundance by a factor of  30 in our standard model.
\citet{Geers2009} find after an analysis of protoplanetary disc spectra a depletion factor of PAHs of  10--20 compared to the ISM.
 \citet{Maaskant2014} also reduce their PAH mass to obtain a PAH-to-dust mass ratio of $5 \cdot 10^{-4}$ (corresponding to 1/5 ISM PAH abundance) in order to match their radiative transfer model to the observed disc spectrum of HD\,97048.
These results might indicate that those discs experienced no cycles or only a few cycles of coagulation so that most PAH were preserved in the gas-phase.
In contrast, discs without detectable PAH emission might have experienced more cycles of PAH processing so that most PAHs in these discs are  frozen out on grains and are unable to enter the gas-phase again.

%%%%%%%%%%%%%%%%%%%%%%%%%%%%%%%%%%%%%%%%%%%%%%%%%%%%%%%%%%%%%%%%%%%%%%%%%%%%%%
%%%%%%%%%%%%%%%%%%%%%%%%%%%%%%%%%%%%%%%%%%%%%%%%%%%%%%%%%%%%%%%%%%%%%%%%%%%%%%
%%%%%%%%%%%%%%%%%%%%%%%%%%%%%%%%%%%%%%%%%%%%%%%%%%%%%%%%%%%%%%%%%%%%%%%%%%%%%%

\subsection{Dust population and vertical mixing}
Our evolution model is parametrised by $\sigma_\text{PAH}/\sigma_\text{dust}$ making it possible to adopt it to other dust populations than our standard model (STD).
Therefore, the dust population is a free parameter that allows for a variation in the depletion of the gas-phase PAHs.
The first PAH processing cycle is especially important for the evolution of the PAHs as the largest fraction is lost if $\sigma_\text{PAH}/\sigma_\text{dust}$ is limited by the maximum recoverable fraction.
As this value is achieved with the ISM abundance of PAHs, the PAH depletion is strongly dependent on the collisional cross-section $\sigma_\text{dust}$ of the dust population.
More precisely, in the case of rapid coagulation ($\sigma_\text{PAH}/\sigma_\text{dust} > 30$),  the PAH loss during the first cycle is proportional to $\sigma_\text{dust}$ as the abundance of PAHs after one cycle is independent of the initial $\sigma_\text{PAH}/\sigma_\text{dust}$ ratio.\\
\\As a consequence, two similar discs with similar host stars but different dust populations will have a different depletion of PAHs.
The larger the dust grains and smaller the total grain population cross-section, the more PAHs will stay frozen out on the dust and the larger the depletion.
For this purpose, we introduce three other dust populations: a model with only small grains (SGs), a model with dominantly large grains (LGs), and a model with very large grains (VLGs).
For each of them we can calculate the initial $\sigma_\text{PAH}/\sigma_\text{dust}$ ratio and then the PAH depletion by one cycle of coagulation, adsorption, and desorption.
Using the initial $\sigma_\text{PAH}/\sigma_\text{dust}$ ratios in Table \ref{tab:disc_models}
we find a PAH depletion factor of $\approx$ 300 for the standard model (STD), a depletion factor of $\approx 2$  in the small grain model (SG), a depletion factor of $\approx$ 20\,000 for the large grain model (LG), and an extreme depletion of $\approx$ 100\,000 for the very large grain population (VLG).\\
\\A Herbig star disc with a sufficient UV field but no visible PAH signatures must  therefore   have depleted its PAHs in most parts of the disc.
For the inner part (<50\,au), the disc must have large grains present where PAHs coagulate. Furthermore, if the disc is weakly turbulent ($\alpha \leq 10^{-4}$), the disc must be sufficiently old to have enough time to allow for at least a few PAH processing cycles.
If instead the disc is intermediate or strongly turbulent ($\alpha > 10^{-4}$), the disc can be young as a PAH processing cycle only takes 1\,Myr. 
Finally it must destroy its PAHs or not have any PAHs present at all.
However for the depletion of PAHs in the outer disc regions, we require from our model that the processing of PAHs occurs before the disc stage, such as the infall; that there is a dense disc structure that induces $G_0 \approx 1$ regions to also allow for adsorption; and that the PAHs are destroyed or that there are no PAHs at all.
Furthermore, other effects that are beyond the scope of this work, such as disc structures, partial shadowing of the disc, inclination of the disc, or dust evolution will likely affect the observability of PAH features as well.

%%%%%%%%%%%%%%%%%%%%%%%%%%%%%%%%%%%%%%%%%%%%%%%%%%%%%%%%%%%%%%%%%%%%%%%%%%%%%%
%%%%%%%%%%%%%%%%%%%%%%%%%%%%%%%%%%%%%%%%%%%%%%%%%%%%%%%%%%%%%%%%%%%%%%%%%%%%%%
%%%%%%%%%%%%%%%%%%%%%%%%%%%%%%%%%%%%%%%%%%%%%%%%%%%%%%%%%%%%%%%%%%%%%%%%%%%%%%

\subsection{PAH desorption through high-velocity grain collisions}
In this section we   discuss the importance of dust grain collisions for the PAH evolution.
So far, we have only considered desorption by photon-induced processes and neglected other desorption mechanisms.
Therefore, we want to consider the relevance of mechanical processes such as the abrasion of PAHs by dust grain collisions.
For this to be relevant, there must be a sufficient number of  high-velocity impacts that mechanically erode the surface of the dust grains. 
We estimate the minimum velocity for these events by considering the necessary collision velocity required for the abrasion of graphitic surfaces.
\citet{Jones1996} reports for graphitic surfaces that the minimum critical impact velocity for the occurrence of splinters is $\approx$1\,km/s.
These required velocities are supported by the laboratory studies of \cite{Zamith2020}, who report a required collisional energy of 0.7\,eV - 1\,eV (0.8\,km/s - 1\,km/s) for the collisional dissociation of pyrene clusters.
Such high velocities cannot be achieved by Brownian motion of dust grains because extreme temperatures are required.
However, turbulent collisions between dust grains can reach much higher collision velocities than pure thermal motion.\\
\\\citet{Cuzzi2003} and \citet{Ormel2007} give close-form expressions to calculate the relative particle velocities between two particles with different Stokes numbers in a turbulent disc.
In their model the highest relative velocities are reached when a particle with a Stokes number $\text{St} = t_\text{stop}/t_\text{edd} = 1$ (the ratio of the particle stopping time to the eddy turnover time) is involved in a collision.
Then the relative velocity is comparable to the local turbulent velocity of the gas:
\begin{equation}
    \Delta v \approx v_\text{g} = \sqrt{\alpha} c_\text{s}\text{.}
\end{equation}
This means that even for an 1000\,K gas an unrealistically high turbulence parameter of $\alpha \approx1$ is required to achieve collision speeds of 1\,km/s.
These conditions are typically not met in protoplanetary discs despite the lack of St $=1$ grains close to the photosphere.
Therefore, we consider the mechanical removal negligible for the general evolution of adsorbed PAHs on dust grains.

\subsection{Heterogeneous PAH clusters}
One of our simplifying assumptions is that initially all PAHs are present as one species, and therefore we do not consider heterogenous clusters made from different PAH species.
In reality, PAHs will be present as a mixture of different species, possibly a selection of the grandPAHs, a few stable and compact PAH molecules that seem to dominate the astronomical PAHs \citep{Tielens2013,Boersma2015}, as extracted from photodissociation region (PDR) spectra in the ISM.
The heat capacity of a PAH cluster only depends on the number of carbon atoms; therefore, the temperature fluctuations of a mixed cluster will be very similar to a homogenous cluster with the same number of carbon atoms.
However, the size of the grain-touching PAH that binds to the dust grains plays an important role as the binding energy of the PAH species determines the detachability of the overall cluster.
A cluster with a small PAH binding to the grain can be detachable, while the same cluster with a large PAH connecting to the grain might not be detachable.
Thus, the number of carbon atoms in a heterogenous cluster is not sufficient to characterise the cluster and determine its chance of evaporation.
In the case that a small PAH is binding to the grain, more large PAH species can be recovered than estimated through the homogenous model.
The whole cluster will desorb and then the large PAHs will dissociate from the cluster as dissociation can    also be achieved through multi-photon events.
In the contrasting case that a large PAH is binding to the grain, then the cluster will be very unlikely to desorb.
However, as smaller PAHs in the cluster have a lower binding energy to the cluster than the surface PAH to the grain, evaporation of small PAHs can be possible leaving only the medium to large PAH species in the adsorbed cluster behind.  
We therefore estimate that our observed size selection effect for homogenous clusters favouring small PAHs for protoplanetary disc conditions is also present if heterogenous clusters are considered, but it is  not as effective.\\
\\In this regard we note that heterogenous clusters are able to rearrange their molecules.
\citet{Bowal2019}   use molecular dynamics (MD)  to study the morphology of heterogeneous clusters at high temperatures (800\,K and 1600\,K), where the individual monomers can rearrange within the cluster.
Regardless of the initial arrangement of the heterogeneous clusters, the authors find that, on average, the larger PAHs are found closer to the cluster core than the smaller PAHs, which preferably accumulate  on the surface of the cluster.
However, the clusters in these studies are significantly larger than the clusters we can detach in our models (16 monomers).
In addition, the clusters we simulated have a much lower equilibrium temperature, so that rearrangement within the cluster can only take place partially under adsorption of UV photons.
Separate MD studies of gas-phase clusters and adsorbed clusters are needed to clarify whether systematic rearrangements of small PAHs towards the surface are possible and relevant.

%%%%%%%%%%%%%%%%%%%%%%%%%%%%%%%%%%%%%%%%%%%%%%%%%%%%%%%%%%%%%%%%%%%%%%%%%%%%%%
%%%%%%%%%%%%%%%%%%%%%%%%%%%%%%%%%%%%%%%%%%%%%%%%%%%%%%%%%%%%%%%%%%%%%%%%%%%%%%
%%%%%%%%%%%%%%%%%%%%%%%%%%%%%%%%%%%%%%%%%%%%%%%%%%%%%%%%%%%%%%%%%%%%%%%%%%%%%%
%%%%%%%%%%%%%%%%%%%%%%%%%%%%%%%%%%%%%%%%%%%%%%%%%%%%%%%%%%%%%%%%%%%%%%%%%%%%%%
%%%%%%%%%%%%%%%%%%%%%%%%%%%%%%%%%%%%%%%%%%%%%%%%%%%%%%%%%%%%%%%%%%%%%%%%%%%%%%

\section{Conclusions}
We have modelled the coagulation of PAHs and their freeze-out on dust grains in a protoplanetary disc environment around a typical Herbig Ae/Be star.
We find that under disc conditions coagulation is very fast.
Considering only PAHs, clusters can grow to sizes with more than $10^5$ cluster members within a year.
However, in the presence of dust grains, cluster growth is suppressed as clusters start to freeze out on grains.
In this set-up, all PAHs adsorb on a timescale of a year, but the largest cluster sizes that can grow are a factor of  10-100 smaller than in the case without dust grains.
Modelling the UV-driven desorption of these clusters, we find that a critical cluster size exists where PAH clusters cannot desorb any longer.
The larger the cluster building PAH species, the smaller the critical cluster size.
This number is on the order of 10-20 cluster members for small PAHs, such as coronene (C$_{24}$H$_{12}$), and decreases until monomers as large as circumovalene (C$_{66}$H$_{20}$) cannot be desorbed at all through UV photons.
Based on these results, we expect a PAH depletion of at least a factor of  10 in disc environments compared to the interstellar medium if only one cycle of coagulation, adsorption, desorption, and dissociation can happen either during the disc stage or the infall stage.
Given that larger clusters of smaller PAH species can desorb more easily compared to clusters made from large PAH species, we expect a size selection effect with which an initial PAH species distribution is expected to shift to smaller PAHs with every cycle.\\
\\Furthermore, we have used these results in a vertical mixing model where the turbulent turnover time of the largest eddies is the dynamical time.
With a radiative transfer model, we determined that the inner disc ($<$50\,au) provides enough UV shielding ($G_0$) so that PAH clusters can freeze out on dust grains.
However, the outer disc does not provide enough shielding to meet these conditions so that a continuous processing of PAHs is unlikely as PAHs simply cannot adsorb on dust grains.
We find that through turbulent processing, the inner disc can have multiple coagulation--adsorption cycles over the lifetime of the disc.
There the depletion of PAHs depends on the turbulent parameter $\alpha$, but also on the PAH species and age.
However, we find that a weak mixing disc $\alpha < 10^{-3}$ is barely able to process PAHs in less than 5\,Myr.
Therefore, we  expect a similar PAH abundance in the inner and outer disc given that the disc is sufficiently young ($1$\,Myr < $t$  < $5$\,Myr).\\
\\Nevertheless, the abundance difference depends on the initial PAH content of the disc stage and if the PAHs have been processed in the infall stage to adsorbed clusters already.
The James Webb Space Telescope with its instruments NIRcam and MIRI will provide crucial information to further investigate the life of PAHs in protoplanetary discs.

\begin{acknowledgements}
The authors thank the referee Ralf Siebenmorgen for his comments to improve the quality and comprehension of the manuscript. K.L. acknowledges funding from the Nederlandse Onderzoekschool Voor Astronomie (NOVA) project number R.2320.0130. C.D. acknowledges funding from the Netherlands Organisation for Scientific Research (NWO) TOP-1 grant as part of the research program “Herbig Ae/Be stars, Rosetta stones for understanding the formation of planetary systems”, project number 614.001.552. 
Studies of interstellar PAHs at Leiden Observatory are supported through a Spinoza award from the Dutch research council, NWO.
\end{acknowledgements}

%%%%%%%%%%%%%%%%%%%%%%%%%%%%%%%%%%%%%%%%%%%%%%%%%%%%%%%%%%%%%%%%%%%%%%%%%%%%%%
\bibliographystyle{aa} % style aa.bst
\bibliography{library.bib} 
\newpage
%%%%%%%%%%%%%%%%%%%%%%%%%%%%%%%%%%%%%%%%%%%%%%%%%%%%%%%%%%%%%%%%%%%%%%%%%%%%%%

\begin{appendix}

\section{Calculation of the cluster size}
\label{app:cluster_size}
\begin{figure}[h!]
    \centering
    \includegraphics[width=\linewidth]{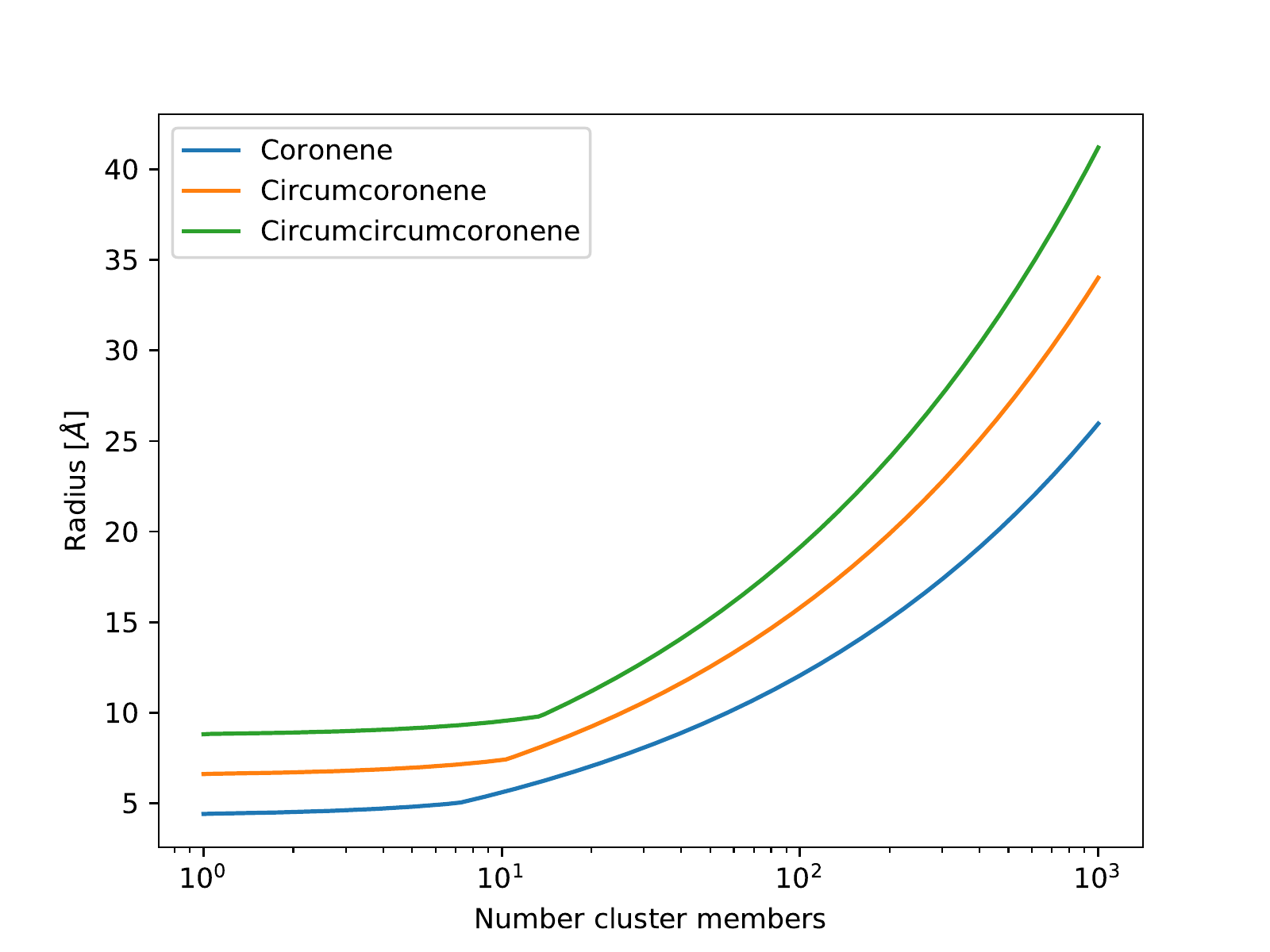}
    \caption{PAH radii given the model in Appendix \ref{app:cluster_size}. For large clusters, the clusters grow in three dimensions, like a dust grain. Assuming PAHs grow initially in stacks, we linearly interpolate between the monomer and the cluster size where a stack has similar width and length.}
    \label{fig:PAH_radii}
\end{figure}
Given the growth cases in section \ref{sec:cluster_size}, we need to calculate the collisional cross-section of all cluster sizes.
Therefore, we define the effective cluster radius as
\begin{equation}
    r =
    \begin{cases}
        r_0 =  0.9 \text{\AA} \sqrt{N_\text{C,0}} &\text{if $N=1$}\\
        r_1 =  0.9 \text{\AA} \left((N_\text{C}-N_\text{C,0})\frac{r_2(mN_\text{C,0})-r_0}{mN_\text{C}-N_\text{C,0}}+r_0\right) &\text{if 1 < N < m}\\
        r_2 = 0.9 \text{\AA} \left(N_\text{C}\right)^{1/3} &\text{if $N\geq m$}
    \end{cases}
,\end{equation}
where $N$ is the number of molecules in the cluster. The parameter
$r_0$ defines the size of a monomer, which is proportional to the carbon atoms in the PAH $\sqrt{N_\text{C,0}}$ since a PAH monomer grows in a plane.
We expect the largest cluster to grow as dust grains, in three dimensions without structure, therefore $r \propto N_\text{C}^{1/3}$.
In our model this occurs once the length of a stack \citep{Rapacioli2005} of PAHs   exceeds the diameter of a monomer.
This occurs at a cluster with $m$ monomers:
\begin{equation}
    m = \frac{2 \cdot 0.9 \text{\AA} \sqrt{N_\text{C,0}}}{1.42 \text{\AA}} +1
.\end{equation}
Here  we   assume that the distance between two stack-bound monomers is similar to the interlayer-distance of graphene (1.42\AA).
Between the monomer and grain-like case, PAHs will likely grow as single stacks, even though multi-stack structures are also allowed depending on the temperature of the cluster.
As this introduces an orientation dependent collisional cross-section, we choose to simplify this problem by linear interpolation between $r_0$ and $r_2$.

\section{Test case for dust density distribution}
\label{app:dust_density}
\begin{figure}
    \centering
    \includegraphics[width=0.99\linewidth]{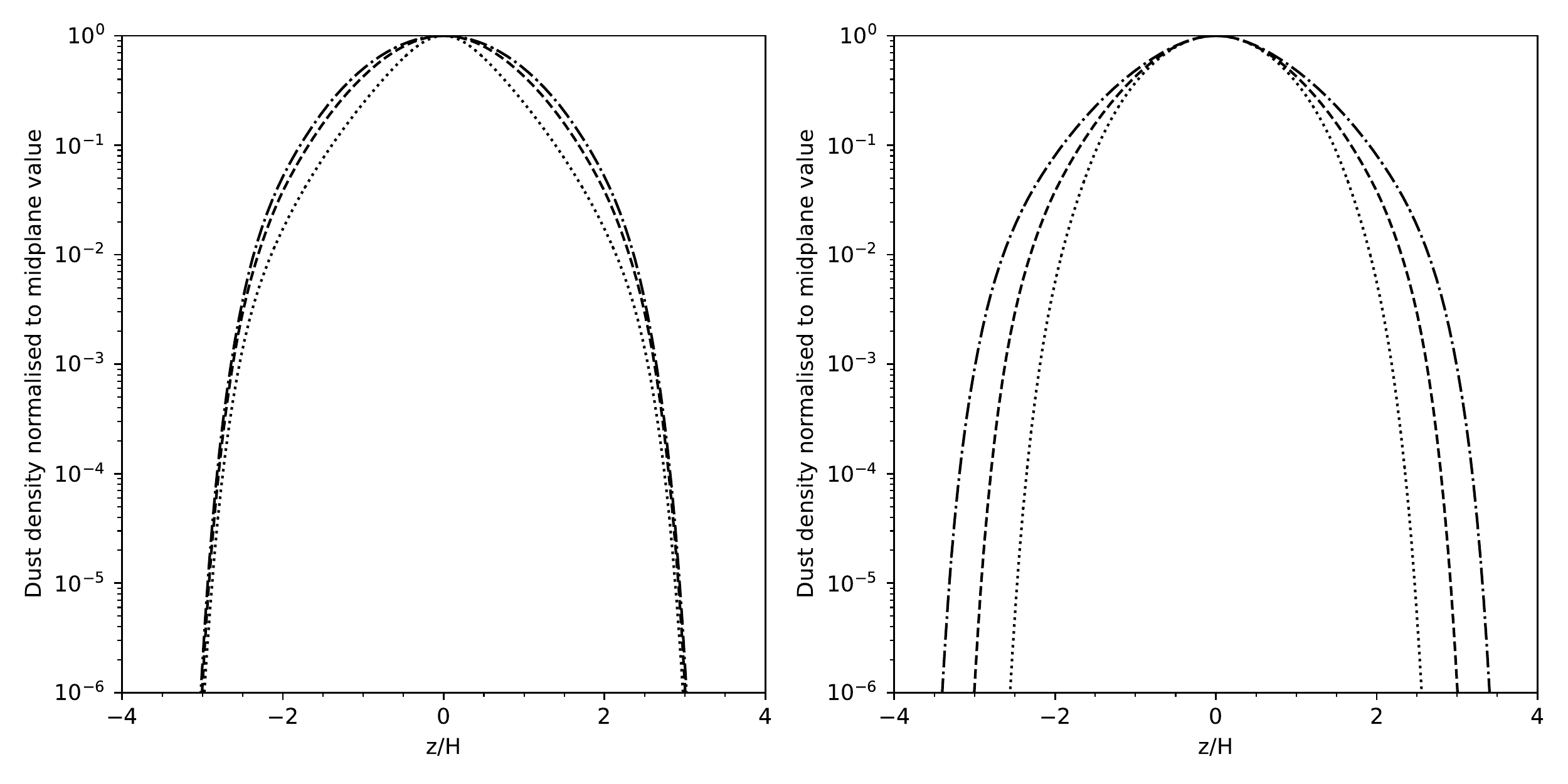}
    \caption{Diagnostic plot to test the vertical grain distribution (same  as figure 13 in \citealt{Fromang2007}). \textit{Left}: Variation of the midplane velocity fluctuations. \textit{Dotted}: $\delta v_\text{z,mid}/c_\text{s}=0.025$, \textit{dashed}: $\delta v_\text{z,mid}/c_\text{s} = 0.05$, \textit{dash-dotted}: $\delta v_\text{z,mid}/c_\text{s}=0.075$. \textit{Right}: Variation of the upper layer velocity fluctuations. \textit{Dotted}: $\delta v_\text{z,up}/c_\text{s}=0.075$, \textit{dashed}: $\delta v_\text{z,up}/c_\text{s} = 0.15$, \textit{dash-dotted}: $\delta v_\text{z,up}/c_\text{s}=0.3$.}
    \label{fig:FNcomparison}
\end{figure}

\label{app:desorp}

In order to test the vertical dust density distribution we want to compare our implementation with figure 13 in \citet{Fromang2007}. In their toy model, they use the following coding parameters: gravitational parameter $GM=1$, sound speed at $R_0$ is $c_0=1$, midplane gas density $\rho_0=1$, reference radius $R_0=1$, aspect ratio $H/R=0.1$, and Stokes number in midplane at reference radius $(\Omega \tau_\text{s})_0=0.001$. Without specifying a disc model, this can be achieved by setting the grain size to $a=0.0001$ and the bulk grain density $\rho_\text{s}=1$. Because \citet{Fromang2007} evaluate their disc model not at $R_0$ but average between $3 \leq R \leq 5$, we evaluate the vertical dust grain distribution at $R=4$ rather than averaging. As a consequence, the sound speed $c_\text{s}$, the Kepler frequency $\Omega$, the scale height $H$, and the midplane gas density $\rho$ need to be calculated at $R=4$ according to their disc model. This can be done using $c_\text{s} = c_0\sqrt{R_0/R}$, $\Omega = \sqrt{R^{-3}}$, $H = 0.1 R$, and $\rho = \rho_0 \left(R_0/R\right)^{1.5}$.
These values are used with the given expressions in equations \eqref{eq:dust} -- \eqref{eq:tau} to solve the differential equation \eqref{eq:dust} with a fourth-order Runge-Kutta solver.\\
\\Figure \ref{fig:FNcomparison} shows the resulting density distribution with varied velocity fluctuations as in figure 13 in \citet{Fromang2007}. This figure closely agrees with figure 13.

\section{Desorption rates for additional PAHs}
\begin{figure*}[h]
    \begin{subfigure}
    \centering
    \includegraphics[width=0.49\linewidth]{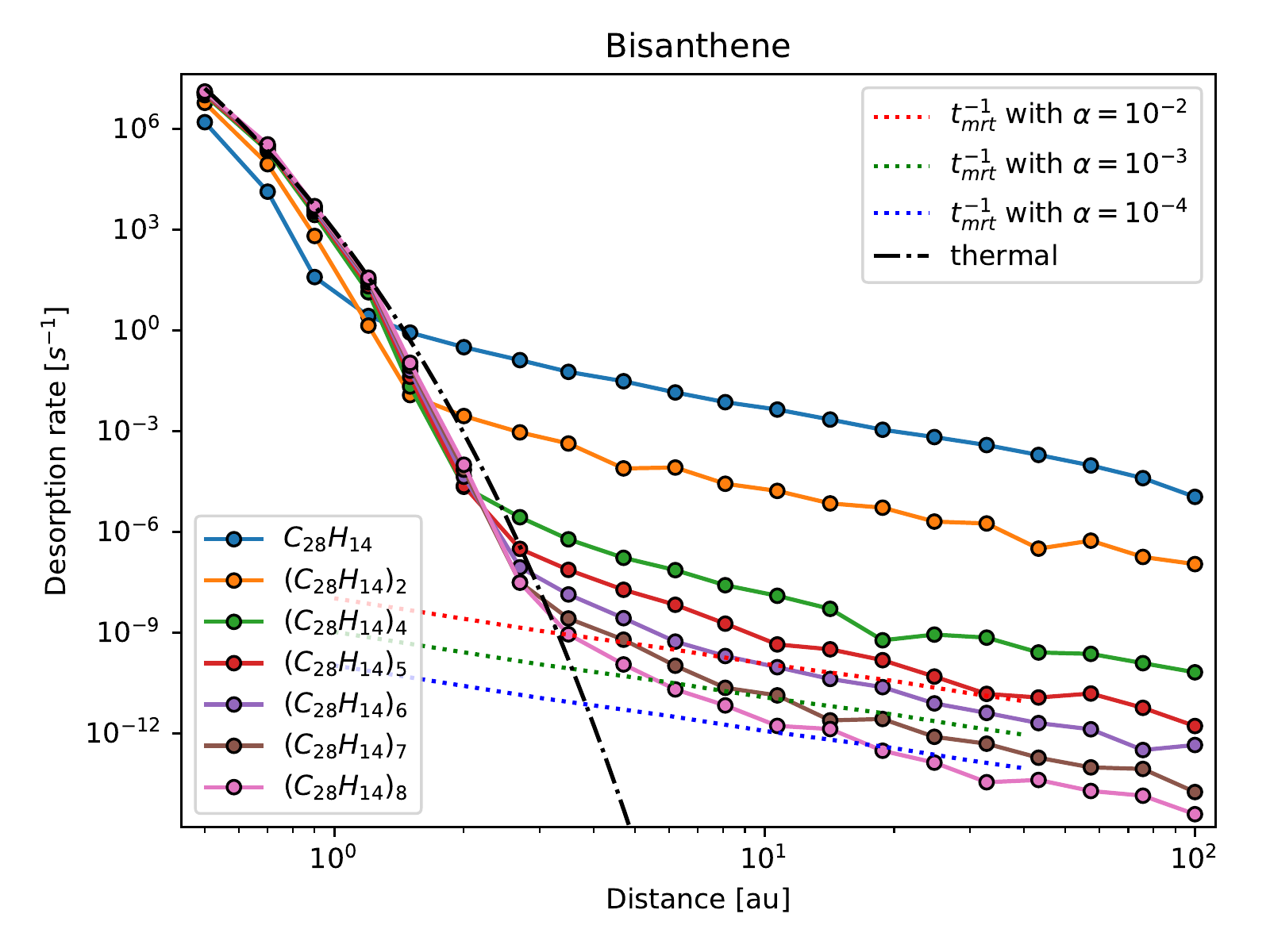}    
    \end{subfigure}
    \begin{subfigure}
    \centering
    \includegraphics[width=0.49\linewidth]{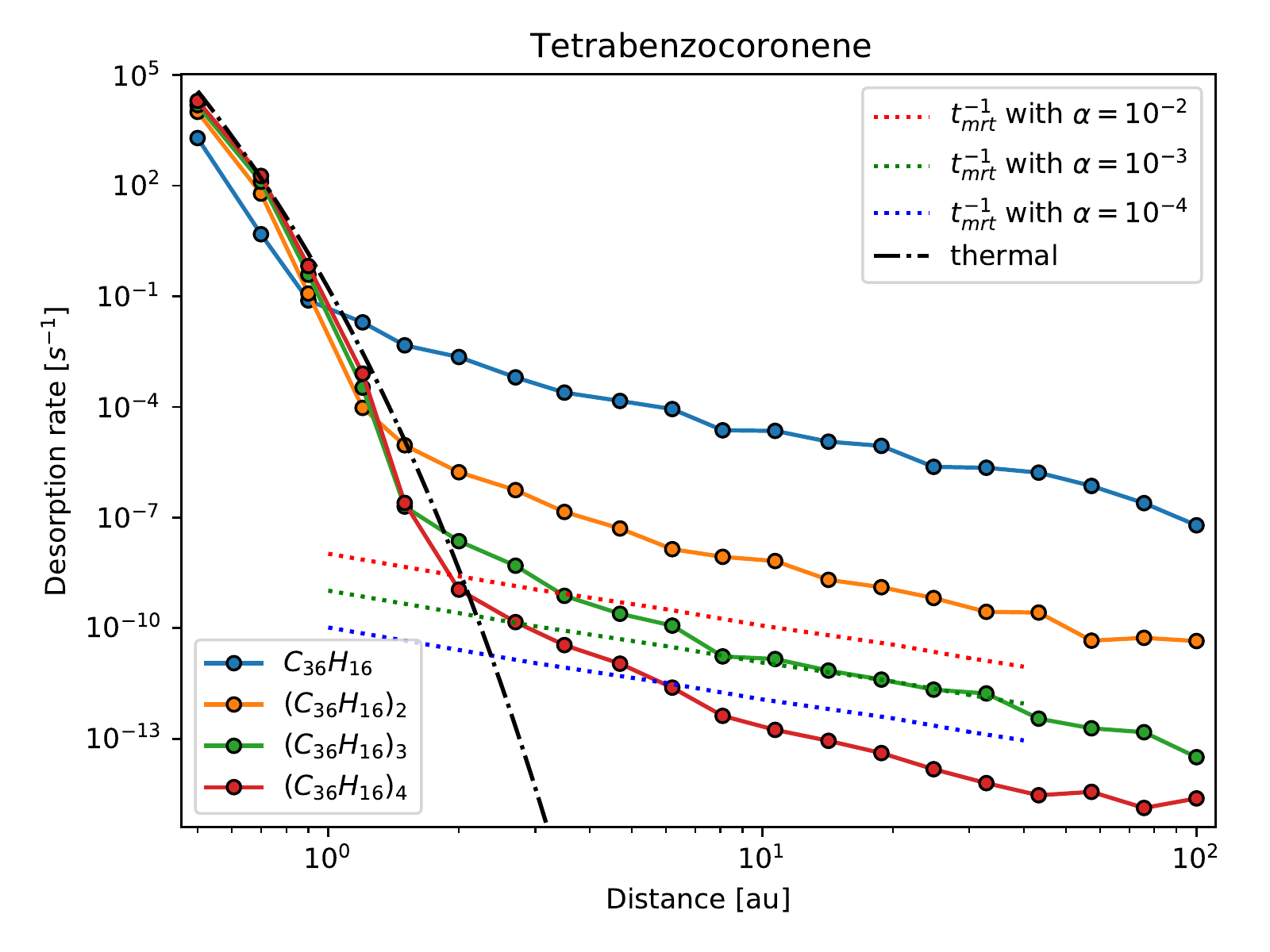}    
    \end{subfigure}\\
    \begin{subfigure}
        \centering
    \includegraphics[width=0.49\linewidth]{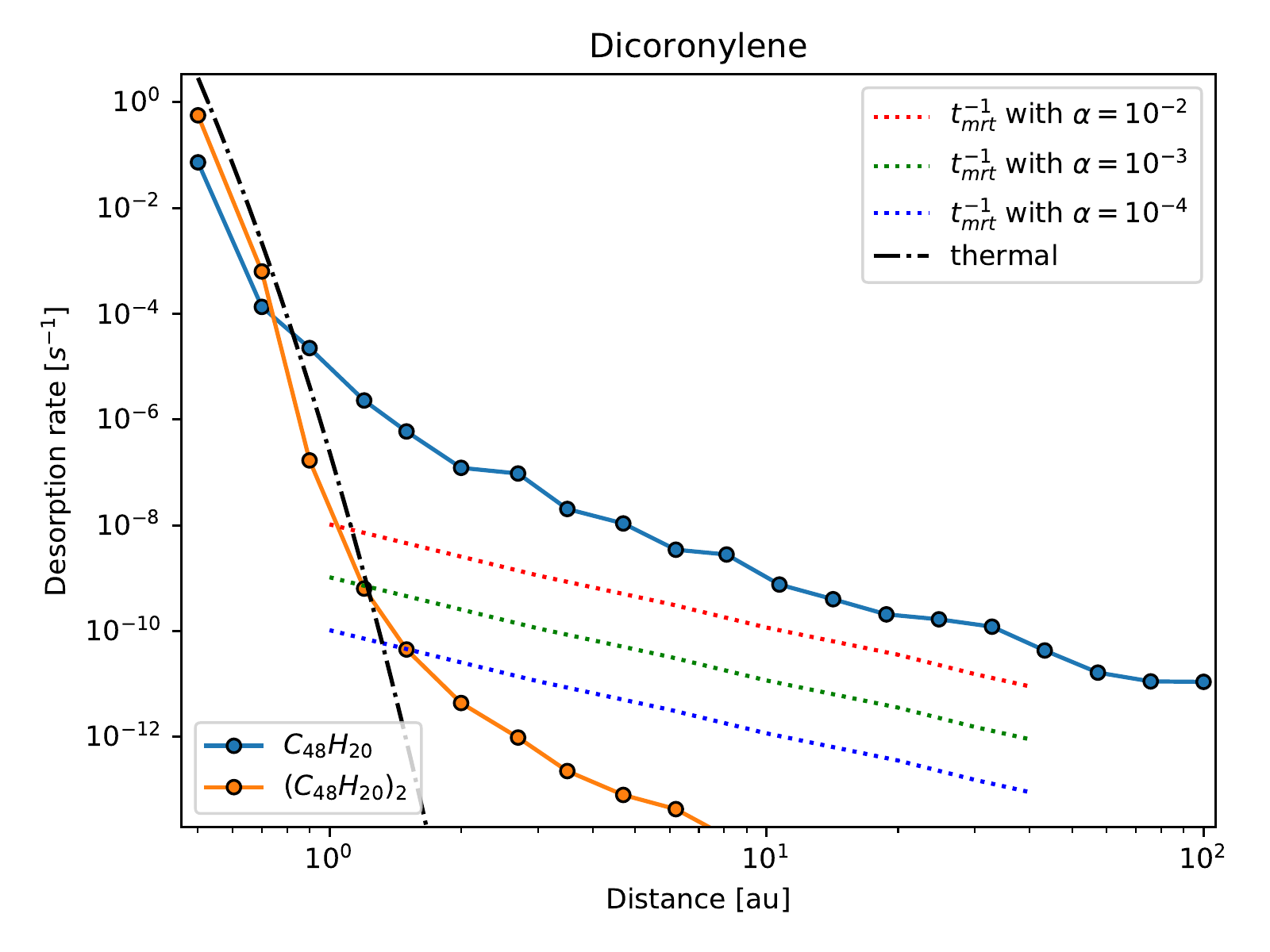}    
    \end{subfigure}
    \begin{subfigure}
        \centering
    \includegraphics[width=0.49\linewidth]{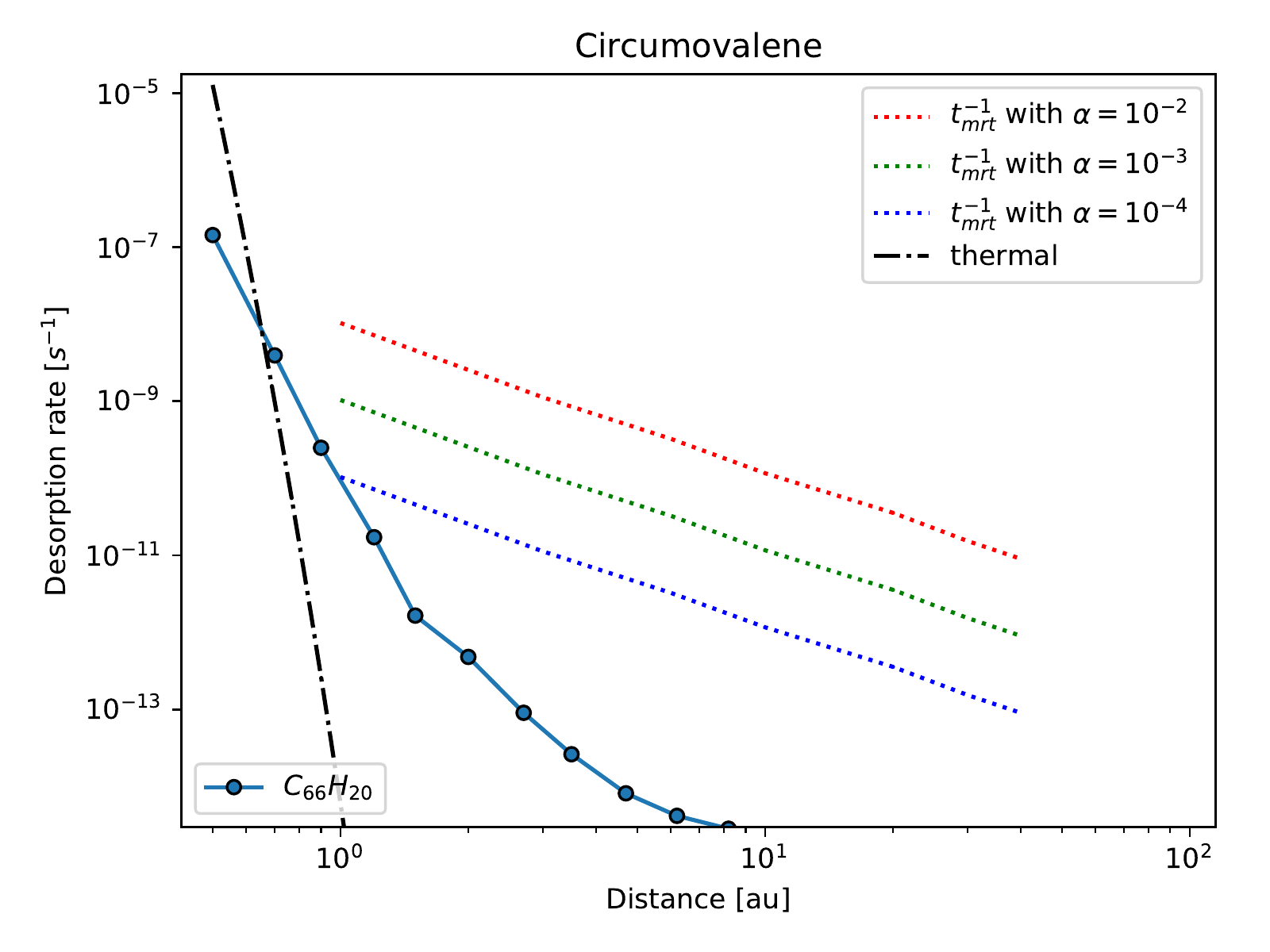}    
    \end{subfigure}
    
    \caption{Desorption rates in the Herbig disc STD photosphere calculated with the Monte Carlo model for increasing cluster sizes, radii, and PAH species. Shown are thermal evaporation rates calculated from eq. \eqref{eq:arrhenius} and eq. \eqref{eq:T_dust} (dash-dotted lines) and   turbulent turn-over rates calculated from eq. \eqref{eq:t_turn} (dotted lines). For increasing PAH size and cluster size, desorption becomes slower until  clusters are unlikely to desorb. PAH coagulation and adsorption on dust grains therefore lead to loss of gas-phase PAHs to the grains.}
    \label{fig:desorb_r_appendix}
\end{figure*}

Figure \ref{fig:desorb_r_appendix} shows additional PAH species for which we have performed desorption calculations.
Bisanthene, tetrabenzocoronene, and dicoronylene are species that are between the smallest and largest shown species by size in the results section. Circumovalene is the smallest PAH species that we cannot desorb through UV photons from the grains any more.

\end{appendix}
\end{document}